\documentclass[pdflatex,sn-mathphys-num]{sn-jnl}%

\usepackage{graphicx}%
\usepackage{multirow}%
\usepackage{longtable}%
\usepackage{siunitx}%
\usepackage{amsmath,amssymb,amsfonts}%
\usepackage{amsthm}%
\usepackage{mathrsfs}%
\usepackage[title]{appendix}%
\usepackage[table]{xcolor}%
\usepackage{xcolor}%
\usepackage{textcomp}%
\usepackage{manyfoot}%
\usepackage{booktabs}%
\usepackage{algorithm}%
\usepackage{algorithmicx}%
\usepackage{algpseudocode}%
\usepackage{listings}%

\newcommand{\red}[1]{\textbf{\textcolor{red}{{#1}}}}

\theoremstyle{thmstyleone}%

\theoremstyle{thmstyletwo}%

\theoremstyle{thmstylethree}%


\begin{document}

\title[Article Title]{Longitudinal Relational Publics and their Discursive Overlap with Issue Publics}%

\author*[1]{\fnm{Alyssa Hasegawa} \sur{Smith}}\email{smith.alyss@northeastern.edu}

\author[2, 3]{\fnm{Judith} \sur{Gilsbach}}

\author[4]{\fnm{Ahana} \sur{Bhattacharya}}
\equalcont{These authors contributed equally to this work.}

\author[5]{\fnm{Holliday} \sur{Sims}}
\equalcont{These authors contributed equally to this work.}

\author[6]{\fnm{Kenneth} \sur{Joseph}}

\affil*[1]{\orgdiv{Department of Mathematics and Computer Science}, \orgname{College of the Holy Cross}}

\affil[2]{\orgdiv{Department of Computational Social Science}, \orgname{GESIS Leibniz Institute for the Social Sciences}}

\affil[3]{\orgdiv{Graduate School of the Social and Behavioural Sciences (GSBS)}, \orgname{University of Konstanz}}

\affil[4]{\orgdiv{Department of Computer Science}, \orgname{Brown University}}

\affil[5]{\orgdiv{Department of Engineering Education}, \orgname{University at Buffalo}}

\affil[6]{\orgdiv{Department of AI and Society}, \orgname{University at Buffalo}}

\abstract{Online discussions of political issues do not always happen in spaces explicitly dedicated to political talk; these discourses may arise within a knitting forum or on a sports fan page. Whatever one's normative view of politics entering these \emph{online third spaces}---contexts that are neither work nor home, where informal interaction unfolds around a shared interest, identity, or (virtual) place---understanding who brings political issues into them, and when, requires studying these spaces at scale, and studying them at scale requires a construct that captures both who is speaking and who is listening, and that holds up over time. Building on Bruns' distinction between participant-centered personal publics and post-centered issue publics, we introduce the \emph{longitudinal relational networked public} (or, simply, the longitudinal public): the coupling of discourse produced by a socially connected set of creators with the durable attention their shared audience gives it. The longitudinal public departs from related relational constructs in three ways: it is anchored in the attention patterns of a non-elite, population-level audience; it treats within-public structure as an object of analysis rather than assuming homogeneity; and it incorporates the audience as a force that shapes creator discourse. We identify 150 longitudinal publics from the following ties of a panel of Twitter/X users matched to U.S. voter records, then measure their discursive overlap with the electoral politics and Black Lives Matter issue publics across 2020, a period spanning the murder of George Floyd and the general election. The spaces that best fit the idea of a third space have the most politically heterogeneous audiences---and thus, potentially, the most room for cross-partisan talk. Overlap varies systematically with public type, with audience age and partisanship, and with creator centrality: the creators most central to a public engage least with political issue publics, a pattern strongest in publics built around local news. These results show how a relational, audience-aware construct can reveal where and through whom political talk enters everyday online life.}


\maketitle
\section{Introduction}\label{sec1}

In June 2019, Ravelry, a well-known knitting pattern website, announced they were banning content supporting Donald Trump \citep{pageHowBanProTrump}. Conservative knitters protested the ban and created new online spaces where pro-Trump content was welcome; some liberal-leaning knitters were also uncomfortable with the free speech implications of the ban \citep{pageHowBanProTrump}. \citet{bickel-knitting-2020} studied the reactions posted, whether supporting or protesting the ban, among a long-standing community of yarn dyers on Instagram. Ultimately, \citet{bickel-knitting-2020} concluded that the dyers' shared social context and specific domain knowledge allowed them to speak confidently about the moral aspects of this political controversy. 

This group of yarn dyers is an example of an \emph{online third space}: a context that is neither workplace nor home, where informal interaction unfolds around a shared interest and political talk can bleed into ordinary conversation \cite{talk2015third}---about yarn, or sports \cite{towler2020shut}, or funny videos \cite{rajadesinganPoliticalDiscussionAbundant2021}.  Unlike the spaces generally studied by researchers observing political discourse on social media, third spaces are not explicitly political in focus \cite{talk2015third}. Understanding political discourse in third spaces matters, however, because informal political discourse in these everyday online spaces not only has the potential to bridge the personal and the political for non-elite social media users, it also may drive cross-partisan political talk due to their relative lack of \emph{a priori} partisan sorting and relative abundance of weak ties with cross-partisan alters \cite{talk2015third}. At the same time, online third spaces exist within a much broader online ecosystem of ``participants and posts'' \cite{brunsPublicSphereNetwork2023} where political discourse emerges. 

Studying how political discourse emerges (or does not emerge) within specific subspaces of this ecosystem raises two challenges. First, we must define what should ``count'' as a well-defined subspace (e.g. a particular third space). If we wish to do so retrospectively and at scale, we likely must make use, at least in part, of large-scale digital trace data. Given the scale of this data, we must further be selective about where we look for political discourse, working to most effectively ``glean that which is beginning to percolate politically'' \cite{dahlgren2006doing} in these spaces by focusing on moments of systemic shock. At the same time, characterizing shocks' impacts on the prevalence of political discussion requires knowledge of a space's baseline structure and activity. Quantifying this baseline means knowing what binds this group together, what its levels of political discourse look like day to day, who speaks, and who is listening. 

To this end, the goals of the present work are first to advance, both theoretically and methodologically, our ability to define and detect distinct online third and political spaces, and then reason about their role in audience members' civic engagement. We then provide a case study to advance substantive understanding of where and when political discussion might emerge in online third and political spaces. Theoretically, we understand third spaces as one of many kinds of online networked publics \cite{habermas_structural_1991,boyd_social_2010,boeder_habermas_2005,warnerPublicsCounterpublics2021,aslama_public_2009}. Although online publics revolve around both participants and posts \cite{warnerPublicsCounterpublics2021}, \citet{brunsPublicSphereNetwork2023} notes that the dominant means of \emph{operationalizing} them typically center on either one or the other, producing distinct theoretical constructs and empirical analyses. Building on \citet{schmidtTwitterRisePersonal2014}, Bruns conceptualizes participant-centered publics as \emph{personal publics}---“the sum of all such exchanges surrounding” a given actor (p. 72). In contrast, post-centered publics emerge when users engage collectively around a shared issue, interest, or world event, yielding \emph{issue publics} \cite{brunsPublicSphereNetwork2023}. Consistent with this conceptual split, empirical studies generally opt either to detect issue publics using content signals \cite[e.g. hashtags, see for example][]{jackson_ferguson_2016, gallagher_reclaiming_2019, shugars_pandemics_2021, clark_white_2019}, or to study personal publics identified using relational network ties \cite[e.g.][]{zhang_social_2021}. %

The online spaces of interest in the present work are, in Bruns' framework, a \emph{collection of} personal publics that are \emph{overlapping in their audiences} and \emph{stable over time}. Bruns alludes to the importance of such stable, overlapping collections of personal publics, but does not name them. We therefore directly extend Bruns' work by introducing the idea of a \emph{longitudinal relational networked public}, or more simply, a \emph{longitudinal public}. A \emph{longitudinal public} defines the coupling of the discourse produced by a socially connected set of creators and the durable attention given to that discourse by the creators' shared audience. Staying within Bruns' framework, we understand shocks that introduce politics into these spaces as inducing \emph{discursive overlap} between a given longitudinal public and an issue public. Issue public discourses, typically identified through content signals, can enter into, recede from, and re-emerge within longitudinal publics---including seemingly apolitical third spaces---over time. We analyze discursive overlap to characterize its heterogeneity across issues, time, and longitudinal publics, in ways that \citet{brunsPublicSphereNetwork2023} notes are of particular interest but rarely studied.

\begin{figure}[!t]
    \centering
    \includegraphics[width=\linewidth]{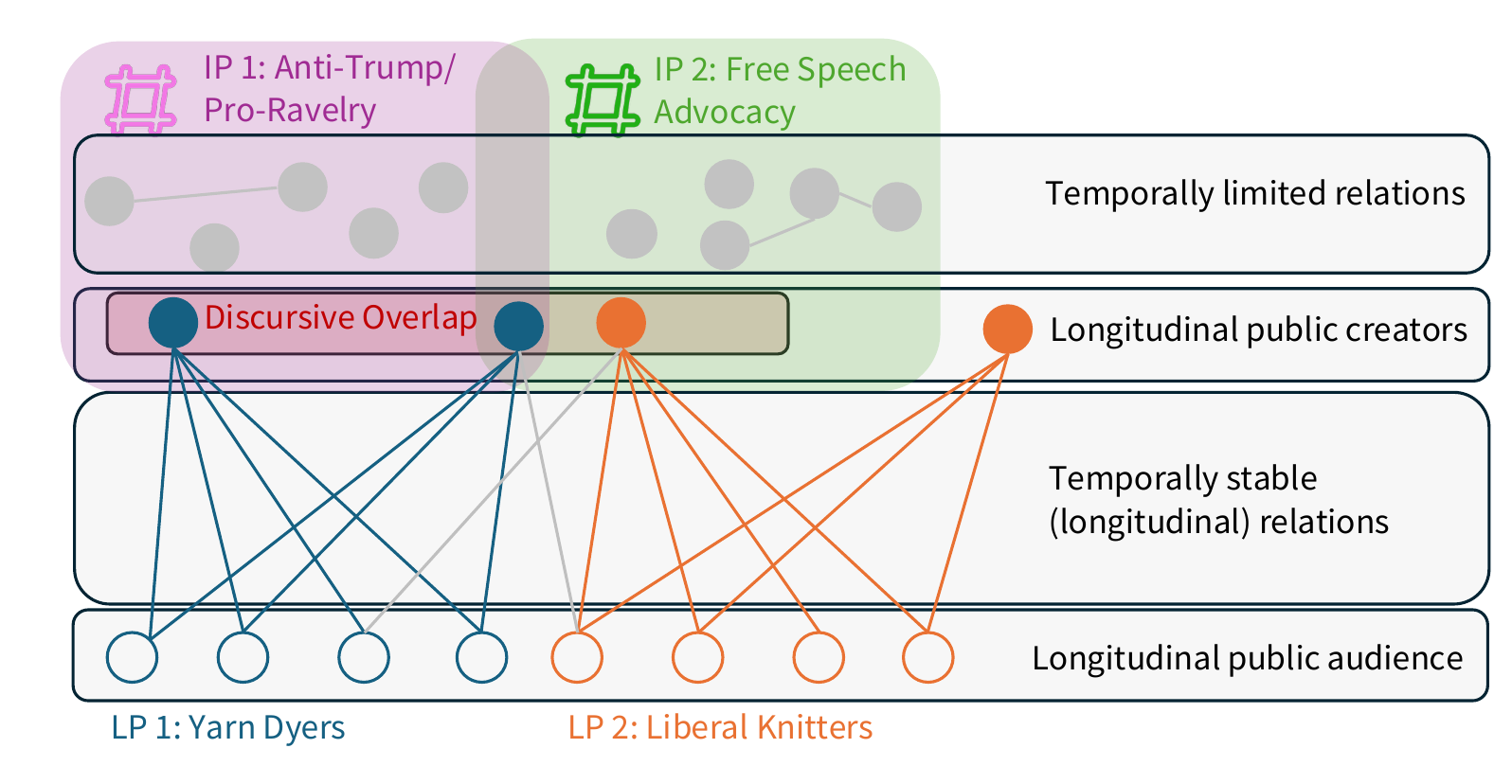}
    \caption{A toy schematic explaining the concept of the longitudinal public. This diagram draws on the Ravelry example to illustrate how longitudinal publics (abbreviated here as LP) overlap with issue publics (abbreviated as IP).}
    \label{fig:longpub}
\end{figure}

Figure~\ref{fig:longpub} provides a toy overview of the concepts of a longitudinal public and their discursive overlap with issue publics in the context of the Ravelry example. The yarn dyers studied by \citet{bickel-knitting-2020} constitute one longitudinal public; for illustrative purposes, we include liberal knitters as a second longitudinal public, drawing on  \citet{pageHowBanProTrump} observation that cohesive groups of liberal knitters coalesced when Trump was elected for the first time. Around the time of the Ravelry ban on pro-Trump patterns, a novel issue public supporting Ravelry's decision emerged, with creators from both longitudinal publics posting content related to that issue public. Additionally, concerns about the ban's implications for free speech might lead some liberal knitters to post content that overlaps with an intermittently present issue public dedicated to freedom of speech. Both issue publics are temporally limited, but the longitudinal publics are relatively stable over time.  

The longitudinal public connects most directly to the \emph{flock} construct introduced by \citet{zhang_social_2021}. Both the longitudinal public and the flock operationalize publics relationally, and moreover leverage the same bipartite clustering approach empirically. However, as we outline in detail in Section~\ref{lp_intro}, the longitudinal public departs from the flock in three important ways: it is anchored in attention patterns of a non-elite, population-level audience rather than attention of elite actors; it treats within-public structure as an object of analysis rather than assuming heterogeneity, and it incorporates the audience, via the ``total communicative output'' framework from \citet{beers_measuring_2025}, as a factor that shapes and is shaped by creators' discourse. To this end, we embrace the inseparability of actors and discourses, explicitly situating patterns of attention and of content production as intertwined within the construct of the longitudinal public \cite{brunsPublicSphereNetwork2023}.

In the following sections, we provide additional details on the longitudinal public and its relation to other theoretical constructs. We then turn to a case study where we identify and validate a set of 150 longitudinal publics on Twitter/X in 2020 using vintage sparse PCA, a straightforward spectral method for clustering bipartite networks that both generalizes and is more computationally efficient than a number of other common approaches (in particular, the degree-corrected stochastic blockmodel) \citep{rohe_vintage_2023}. Having identified the set of longitudinal publics of interest, we turn towards two research questions:
\begin{itemize}
    \item \textbf{RQ1:} What kinds of longitudinal publics existed on Twitter in 2020?
    \item \textbf{RQ2:} What factors are associated with heterogeneity in the discursive overlap between issue publics and longitudinal publics \emph{over time (RQ2a)}, \emph{across (RQ2b)}, and \emph{within (RQ2c)} longitudinal publics?
\end{itemize}

We address RQ1 via qualitative analysis of the identified longitudinal publics, assigning each longitudinal public a label (e.g. ``Wisconsin Media Accounts'' or ``Progressive Liberals''), and developing a higher-level typology as well. Using the longitudinal public construct here allows us to identify third \citep{talk2015third} spaces, as well as other spaces defined by particular identities, geographies, and politics defined by a representative panel of U.S. voters who are also Twitter/X users \citep{hughes_using_2021}, rather than the spaces occupied by some population of elites that would be uncovered using the audience selection method from the flock construct \citep{zhang_assembling_2021}. By examining the spaces that make up the everyday online lives of a target population that is meaningfully defined offline, we can characterize how the political might (and does) bleed into their information diets.

In order to measure the degree of political talk, we analyze the extent to which the content produced by each longitudinal public's creators experiences discursive overlap with issue publics. Our emphasis on third spaces makes the longitudinal public construct particularly well-suited to first detecting, then accounting for, variation in the extent to which specific types of political talk permeates different longitudinal publics. We explore RQ2 in the context of two different, previously studied, issue publics centered around two different issues that were salient in 2020: electoral politics \citep{beers_measuring_2025,shugars_pandemics_2021} and the Black Lives Matter movement \citep{freelon_quantifying_2018,shugars_pandemics_2021}. In the U.S. context, Twitter/X in 2020 was a rich source of public discourse as users weathered upheaval from the COVID-19 pandemic and the 2020 general elections. Misinformation posted on the platform following the 2020 elections helped to incite the January 6th attacks on the U.S. Capitol \citep{starbird_influence_2023}. 

Our dataset consists of tweets collected from the Twitter/X Decahose, which is a 10\% longitudinal random sample of all tweets produced on the platform, created in May 2020 through December 2020. We identify the fractional volume of (re)tweets from members of longitudinal publics for electoral politics and Black Lives Matter-related content using a set of keywords lists from prior work \citep{gitomer_speech_2023,mukerjee_political_2022,shugars_pandemics_2021}. We then use regression analysis to explore how exogenous events (the murder of George Floyd and the 2020 U.S. Presidential election) impact the volume of discursive overlap between longitudinal publics and issue publics, and how this engagement varied over time, across properties of longitudinal publics, and within longitudinal publics, across properties of the personal publics of individual content creators. The longitudinal public construct's awareness of substructures within individual publics and an audience's impact on creator behavior allows us to answer questions like RQ2b and RQ2c: specifically, how do the characteristics of a public's audience correlate with variation in discursive overlaps across publics, and how do individual creators' attributes, along with the traits of a creator's own audience, help explain variation in discursive output across creators within a given public? 

In summary, our work makes the following theoretical and empirical contributions to the literature:
\begin{itemize}
    \item Theoretically, we propose the longitudinal public as a new operationalization of the networked public that helps us conceptualize and measure the discursive overlap over time between stable communities of content creators and their audiences and interest and issue publics. The model is a synthesis of theoretical perspectives from \citet{brunsPublicSphereNetwork2023}, \citet{zhang_social_2021}, and \citet{beers_measuring_2025}.
    \item With respect to RQ1, we use a mixed-methods approach to identify and describe a broad array of longitudinal publics centered on a range of distinct places, identities, and interests. We show that spaces that best align with the idea of an online third space have the most politically heterogenous audiences, emphasizing their potential for cross-cutting political discourse.
    \item With respect to RQ2, we are able to account for heterogeneity over time, within publics, and across publics, at least in part, using attributes of longitudinal publics, their audiences, and their creators. For example, we find that creators who are more central within a longitudinal public, especially those centered around local news, are significantly less likely have discursive overlap with political issue publics.  This advances our understanding of the ways that issue public discourses diffuse into and out of relational publics. 
\end{itemize}

All code and, where possible, data, are available for replication purposes at \url{https://github.com/kennyjoseph/longitudinal_publics}.

\section{(Analyzing) the Longitudinal Public} \label{lp_intro}
While the networked public is itself a useful construct for understanding discourse online, communication scholarship has developed a range of related concepts that enable the interrogation of more specific phenomena. In addition to those outlined by \citet{brunsPublicSphereNetwork2023}, these include affective publics \citep{papacharissi_affective_2012,papacharissi2015affective}, social media local publics \citep{buraiFeelLocalPost2024}, calculated publics \citep{gillespie2014relevance}, and refracted publics \citep{abidin_networked_2021}, among others. Each theoretical construct comes with its own understanding of the discursive and relational structures that underpin a public, its own prerequisites for relevance and inclusion, and its own assumptions about what constitutes a meaningful unit of analysis. \citet{brunsPublicSphereNetwork2023} introduces a taxonomy of the ``primary building blocks of the contemporary communicative environment'' to provide a unifying framework around many of these constructs. \citet{brunsPublicSphereNetwork2023} also calls for work examining the horizontal/vertical interconnectedness of the various kinds of online publics. Our work begins with this call to establish the distinction between relational and issue publics, and aligns it with the idea of a \emph{flock} \citep{zhang_assembling_2021} to respond to this call.

\citet{zhang_social_2021} introduce the concept of the flock in their murmuration framework. A flock is a stable set of actors whose discourse on particular issues represent a useful lens into public opinion. \citet{zhang_social_2021} operationalize flocks by starting from a set of known central opinion leaders, principally sampling from the follower/followee network of these opinion leaders, and then detecting communities (or clusters) within that network. These communities each constitute a flock, a ``[homogenous] collection of similar social media accounts situated in the same neighborhood of a social graph.'' \citet{zhang_assembling_2021} are interested in identifying flocks to act as a representative set of  ``focus group[s]'' for public opinion on an array of topics. Because the social network structure is relatively stable over time, \citet{zhang_social_2021} are able to extract opinions from the sample over a year, noticing discursive changes within flocks in response to relevant events. 

The longitudinal public draws on several important ideas from the flock. First, the longitudinal public, like the flock \citep[and, as we discuss further below, the networked election public][]{beers_measuring_2025} departs from other notions of the public in holding creators as distinct from audiences. Second, similar to the flock, the longitudinal public operationalizes the networked public through stable social relationships online, and incorporates coherent sets of creators that are distinct from their overlapping audiences. Our empirical work even makes use of the same clustering method for bipartite networks of follower relationships. 

Given these similarities, it is reasonable to ask how the longitudinal public actually departs from the flock. Put briefly, we draw on provocations from \citet{brunsPublicSphereNetwork2023} and \citet{beers_measuring_2025} regarding the role of digital ``third spaces,'' the nested nature of networked publics, and the role audiences play in shaping discourse. In doing so, we extend the flock construct into a tool that provides insight into non-elite social media users' information ecosystems and influence on creators. We expand on each of these three points below. 

First, our approach is motivated by a desire to detect spaces that
include digital third spaces \citep{talk2015third}, rather than the
elite networked publics discovered using the flock framework:
\citet{zhang_social_2021} note that they are measuring ``the elite
layer of public opinion;'' given their stated goals, this is ``a
feature, not a limitation.'' The longitudinal public, in contrast,
seeks to understand non-elite users' experiences and opportunities
for political engagement---the (potentially) more mundane online
worlds in which a population that is meaningfully defined offline,
like U.S. voters, at-risk adolescents, or older adults who share
misinformation, is embedded. This conceptual departure drives a
methodological one at what we consider to be the foundation of our
construct: where \citet{zhang_social_2021} construct their audience
using a process designed to discover elite or central actors,
longitudinal publics are defined by the following patterns of an
audience representative of such an offline population---here, a
sample of U.S. voters on Twitter/X that resembles the U.S. voter
base in many ways \citep{hughes_using_2021}. 

The choice of audience
matters. If we adapted the audience construction approach of
\citet{zhang_assembling_2021}, perhaps using the 100 most followed
Twitter/X accounts as seeds for the personalized page-rank algorithm
\citep{chen2020targeted}, we might well discover some non-political
spaces that elites dedicate attention to. However, those spaces are
unlikely to match the ones surfaced by our panel's following ties:
\citet{paul2019elites} found that the network of verified users on
Twitter/X demonstrated higher reciprocity relative to the overall
network, as well as a large number of attracting components,
suggesting that spaces derived from elites' attention patterns will
be more heavily dominated by elites than the third/political spaces
we derive from our panel. \citet{brunsPublicSphereNetwork2023}
points to these everyday spaces, many of which are third spaces, as
occasional yet meaningful contributors to discourse in issue
publics; as part of our case study, we therefore characterize what
kinds of such spaces existed for U.S. voters on Twitter/X in 2020,
with particular focus on discursive overlap with issue publics in
both \emph{a priori} apolitical third spaces \emph{and} politically
focused spaces, in order to understand this opportunity for
informal, everyday engagement with the political.

Second, our approach looks at substructures within longitudinal publics, drawing on the idea from \citet{brunsPublicSphereNetwork2023} that larger, enduring publics like longitudinal publics or flocks are messy, fluid intersections of personal publics that experience temporally limited overlap with issue publics. \citet{zhang_social_2021} analyze flocks, which are composed of collections of content creators, as homogeneous units in terms of both discursive output (``We expect social media public opinion to exhibit homogeneity within a given flock or similar flocks'') and structure (``markedly more followers were shared by members of the same flock than members of different flocks'') \citep{zhang_social_2021}. We adapt the flock construct based on the taxonomy introduced by \citet{brunsPublicSphereNetwork2023}, which leads us to consider what kinds of smaller publics could be nested inside the flock itself. The literature often characterizes public opinion on social media, particularly political opinion, as embedded in ideologically homogeneous sub-networks or communities (see for example \citet{cinelli2021echo}, \citet{colleoni2014echo}, \citet{liTikToksPoliticalLandscape2025}, \citet{gaoEchoChamberEffects2023}). In contrast, RQ2 attempts to explain heterogeneity in discursive overlap across longitudinal publics, but also within them, revealing, for example, that different levels of engagement within longitudinal publics on discourse native to that public predicts variation in the ways in which creators engage with different political issue publics. In broader terms, this emphasis on explaining within-public variation allows us to address key points brought up by \citet{brunsPublicSphereNetwork2023} regarding connections between individual personal publics and the ``serendipity of information flows'' within intertwined networks of personal publics. Among a longitudinal public's creators, who is capable of introducing unexpected yet well-received political talk into the broader discourse? 

Finally, while \citet{zhang_social_2021} analyze output only of creators within a flock, we also look at how characteristics of a longitudinal public's audience shape heterogeneity in discourse and discursive overlap. Creator/audience relationships matter for the health of our information ecosystems: \citet{starbirdUnravelingBigLie2025} found that ``feedback loops'' between creators spreading disinformation and their audiences created a ``fundamentally participatory'' disinformation campaign regarding the 2020 U.S. presidential election's integrity. We therefore draw on the analytical lens from \citet{beers_measuring_2025}, who introduces the networked election public and presents a framework for measuring the ``total communicative output'' of a public as a function of its content creators, their audiences, and platform effects. Specifically, \citet{beers_measuring_2025} expresses the total output of a public, $P$, as 
\begin{equation}\label{eq:beers}
    P = (N_c * A_c) * (R_c) * (N_a * A_a)
\end{equation}
In Equation \ref{eq:beers}, $N_c$ and $A_c$ refer to the number and activity level of creators, respectively. $R_c$ represents the effects of the mediating platform (in this case, Twitter/X), and $N_a$ and $A_a$ refer to audiences' sizes and sharing rates. The multiplicative form encodes the fact that a public's communicative output is dependent on both creators' and audience amplification. We extend this framework here, developing a way of understanding the factors that influence creators' political output and audiences' amplification in longitudinal publics, particularly with respect to characteristics of each longitudinal public and its audience. We then use this extension as the conceptual anchors for the covariates in our regression models. In doing so, we respond to the following from \citet{beers_measuring_2025}:
\begin{quote}
    The use of this model is thus only a simplified starting point for deeper qualitative and experimental research on the complex incentives for change that actors within networked publics face – the how and why to the what of this model's change metrics.
\end{quote}
Our analysis of RQ2 takes up precisely this invitation: by relating creators' discursive overlap to the attributes of their publics and audiences, we begin to supply the ``how and why'' behind changes in a public's political output.

Before turning to the details of our empirical work, we note one final question that might be asked of longitudinal publics: are these patterns of ``participants and posts,'' to return to \citeauthor{brunsPublicSphereNetwork2023}' \citeyear{brunsPublicSphereNetwork2023} framing, actually a \emph{public}? \citet{zhang_social_2021} argue that flocks, at least, are. Does this extend to longitudinal publics? We believe so. Turning beyond \citet{zhang_social_2021} and to \citeauthor{warnerPublicsCounterpublics2021}'s \citeyear{warnerPublicsCounterpublics2021} canonical characterization of a public, we find that the social structures we theorize and study empirically for the most part fit neatly into the qualifications of a public he defines: they are self-organized (followers choose to coalesce around creators), relations among strangers (not all creators or audience members know each other), reflexively circulate discourse over time, and perform (poetic) world-making both within their focal existence and, as we will show, in their intersection with issue publics. 

There is, however, one qualification that merits further discussion: Warner's \citeyear{warnerPublicsCounterpublics2021} emphasis that a public is ``\emph{constituted through mere attention}.'' Here, we conceptualize this attention as occurring through \emph{persistent relational ties}; empirically, we operationalize these ties as \emph{following relationships}. Following is not directly attention – accounts are followed for many reasons beyond a desire to attend to content \citep{barbieriWhoFollowWhy2014}, and many followed accounts send messages that are likely never attended to \citep{grinberg_fake_2019}. However, following relationships are of course \emph{related} to attention. One perspective on this relationship is that at the time point of this study, following relationships represented an “upper bound” on attention, in that users only saw content produced by accounts they were following on their default Twitter feeds. Even today, following relationships remain a critical factor in impacting what users see \citep{duskin2025role}. Moreover, because following is relatively static \citep{zhang_social_2021}, it can shape attention well after the following decision is made \citep{smith2025emergent}. This, in our view, is what is particularly interesting about longitudinal publics: persistent social relationships exist in a form of ``attention ebb and flow'' that is particularly well suited to the kinds of studies that Bruns \citeyear{brunsPublicSphereNetwork2023} seems to find of interest in the overlap between personal and issue publics. Finally, both following and attention are in some sense scarce, in that the breadth of accounts that successive cohorts of new Twitter users follow is shrinking over time \citep{wolf_successive_2022}. The present work makes use of this relationship to provide novel insights in how attention may slowly shift over time in response to events via following relationships.

\section{Methods and Data}\label{methods}

Here, we provide a detailed explanation of our operationalization of the longitudinal public construct we just introduced. It is important to note that this is one specific way to discover longitudinal publics and only represents one way to analyze how issue publics bleed into longitudinal publics: we regard the construct and frameworks of the longitudinal public as our core contribution, and the example we provide should not be seen as the definitive way to discover or analyze longitudinal publics. Rather, we hope it will serve as a jumping-off point for further analysis of longitudinal publics at other times, on other platforms, with other questions in mind. 

\subsection{Identifying Longitudinal Publics}

In order to identify longitudinal publics, we rely on patterns in co-following behavior; that is, we identify longitudinal publics based on clusters of content creators that share similar audiences. The longitudinal publics we study are derived from the following patterns of a panel of roughly 1.2 million Twitter/X users whose accounts have been linked to voter records to provide inferred information about demographic attributes. A longitudinal public's content creators are the most representative users in that public; unless a creator also happens to appear in the audience panel, we cannot make any claims about their demographic attributes. A longitudinal public's shared audience is the set of panel users, for whom we do have inferred demographic information, who follow at least one of its creators. 

\begin{figure}[!t]
    \centering
    \includegraphics[width=\linewidth]{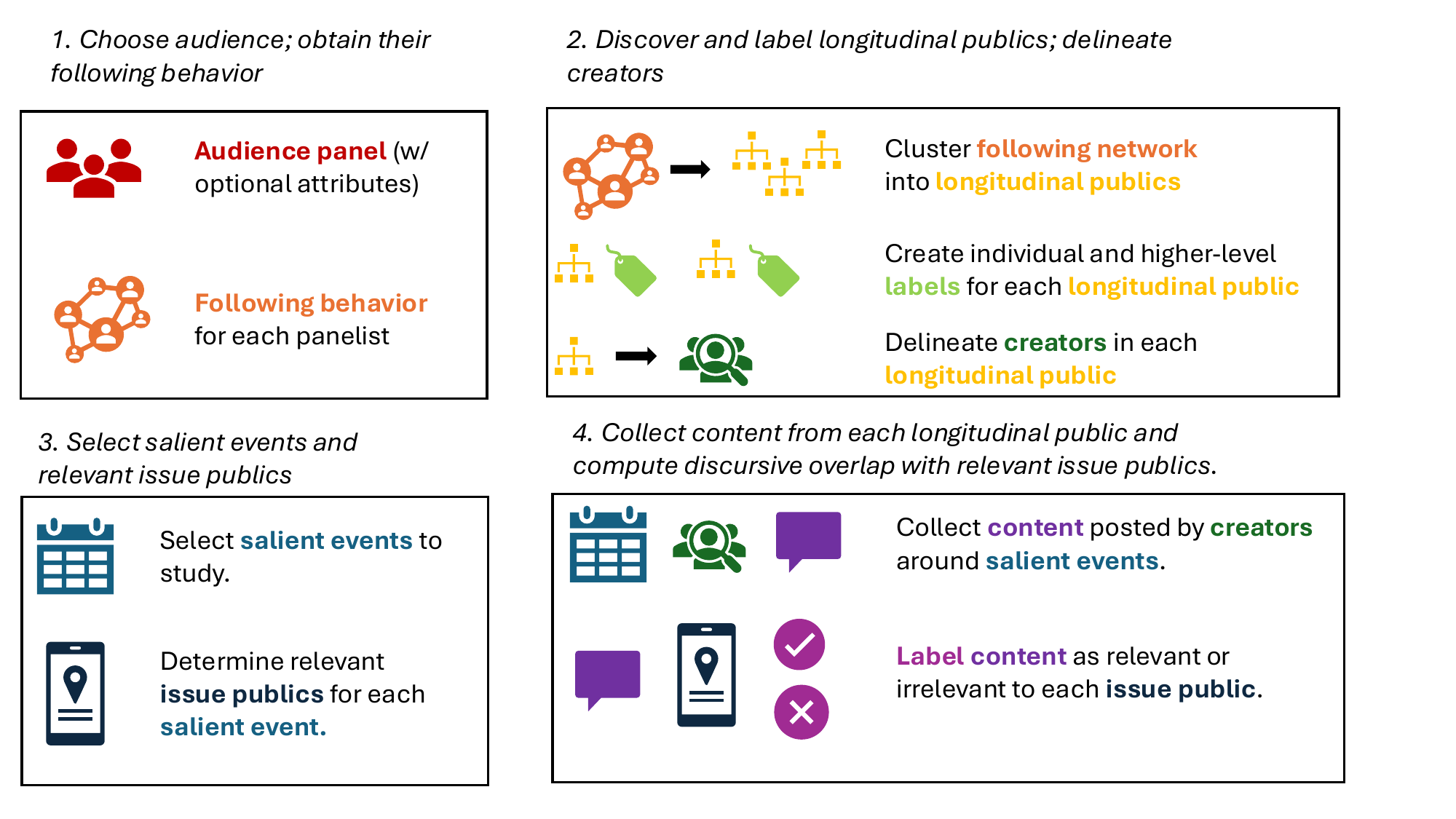}
    \caption{A schematic of the data pipeline for discovering longitudinal publics and determining their discursive overlap with issue publics.}
    \label{fig:pipeline}
\end{figure}

To summarize our approach, we include a high-level schematic of the main workflow for longitudinal public discovery and computation of overlap with issue publics in Figure~\ref{fig:pipeline}.
Immediately below, we review 1) the sample of Twitter users whose following behaviors we study, 2) how we identify patterns in those following relationships, and 3) address ethical concerns regarding our approach. 

\subsubsection{Collection of Following Relationships in a Panel of Twitter/X Users}

To identify longitudinal publics, we study following behaviors in a set of roughly 1.2 million Twitter/X users linked to voter registration records. We refer to this collection of users as our \emph{panel} and to sets of users as panelists or panel members. We begin panel construction with a set of approximately 400M accounts who shared at least one tweet from 2014-2016 that was captured in the Twitter/X Decahose. We then identify users in this sample that can be linked to voter registration records. The methods we use to link Twitter/X accounts to voter records are the same as those described by \citep{grinberg_fake_2019, shugars_pandemics_2021} and extensively validated and compared to samples of American voters obtained using PEW survey methodologies by \citet{hughes_using_2021}. Because we rely on precedent from these prior works, we provide only a brief sketch of the approach used here; we refer the reader to the prior work for additional details. Beginning with the set of users identified in the Decahose, we attempt to match each one to a set of approximately 300M voter records obtained from TargetSmart\footnote{The data was provided under a strict agreement prohibiting us from sharing any identifiable information about the panel.}. We link a Twitter/X account to a voter registration record if each of the following match between the two accounts: first name, last name, city (if present) and/or state. Finally, if there are no other Twitter/X accounts with the given first and last name without a listed location, we add this linked voter record and account to our panel. 

Prior work has estimated that this approach accurately links Twitter/X accounts to roughly 1-3\% of voting-eligible Americans on Twitter/X, and that it does so in ways that are demographically similar to the full population of U.S. Twitter/X users \citep{hughes_using_2021}.  Voter records contain a number of relevant demographics used in this study which we use for our analysis of the audience of influential account clusters; see details below. Following relationships (i.e. who the panel members follow) focuses on a collection of all accounts followed by the panel at a single point in time. This collection draws on data collected over the course of three months from September-November of 2020. In total, the panel followed 68,572,969 unique accounts at that time. In order to perform qualitative analysis on clusters of these followed accounts, we collected user profile data of all 68.5 million of these accounts using the Twitter/X V2 API. This collection occurred over the course of two months, from mid-August through mid-October of 2022. Roughly 15\% of accounts that the panel followed in 2020 were either deleted or suspended by 2022; we still include these accounts in our clustering and note in our qualitative analysis (though not in results we focus on here) where there were prevalent cases of suspended and/or deleted accounts. While the linkage between voter records and Twitter accounts was completed between 2016 and 2017, the tweets we collected, as well as the following relationships, were collected in 2020. 

\subsubsection{Clustering Panel Follower Relationships to Identify Distinct Longitudinal Publics}
 
We first create a matrix where the rows are panel users and the columns are the accounts that they follow. Overall, we have following data for 1,373,158 panel members who follow 68,572,969 distinct accounts, amounting to 603,709,651 following relationships. As with \citet{zhang_social_2021}, we then filter this matrix to exclude low activity panel users and accounts followed by only a few panel members. Specifically, we remove from the columns of the matrix any account followed by fewer than 25 panel members, and subsequently drop any panel user who does not follow at least ten of the remaining accounts.  The filtered matrix consists of 1,226,491 panel users (or 89\% of the original dataset), 2,586,792 followed accounts (3.8\% of followed accounts) and 265,443,277 total following relationships (44\% of the original dataset). 

We then cluster followed accounts based on commonalities in followers; at a high level, followed accounts with many panel followers in common are more likely to end up in the same cluster. The method we use for clustering is the same as \citet{zhang_social_2021}, entitled vintage sparse PCA (VSP) \citep{rohe_vintage_2023}. The method is computationally efficient (we can compute cluster structure in minutes on commodity hardware for networks with tens of millions of links), can provably recover block structure in social networks under fairly loose statistical assumptions, and has strong theoretical connections to known generative processes for bipartite social networks \citep{rohe_vintage_2023}.  Given the tight theoretical connections to our context, the computational efficiency of the method, and the natural connection to the flock approach, we opt for VSP here and do not consider alternatives. While \citet{zhang_social_2021} opt for a model with 100 clusters, we follow \citet{shuster2024supports} and empirically determine the number of clusters by finding a value for $k$ where clusterings are consistent with most other values of $k$, as based on \emph{Average Mutual Information} (AMI) \citep{steuer_mutual_2002}. The setting of $k=150$ minimizes the average AMI of the clustering relative to a range of $k$ between 50 and 225. 

\subsubsection{Ethical Considerations}
Our approach for identifying the panel under study has been approved by the Institutional Review Board (IRB) at Northeastern University. Notably, we do not in our work seek to \emph{infer} the information used for matching (names and locations). Because of this, anyone who distinguishes their information on Twitter/X in any way from what they put in their voter record would be excluded from our sample. In addition to being an important ethical point for our team, this also means that we do not violate the Twitter/X Terms of Service.  

At the same time, there are a number of serious ethical implications of our work that we would be remiss not to identify here. We have provided a detailed analysis of some of these implications in prior work \cite{hughes_using_2021, wojcik_us_2019, grinberg_fake_2019}, but we summarize the main important points here. First, with respect to the voter records themselves, the use of a number of categorization schemes in administrative data can, if used carelessly, be detrimental to computational social science research \citep{spiel2019better}. As such, we aim to use these data only when addressing research questions below for which we have a theoretical reason to believe that dimension is of interest. Additionally, our linking approach limits our analysis to individuals who harbor certain forms of privilege that makes them able to share their true names and locations. This analysis thus may exclude certain marginalized populations; we are therefore careful to caveat our findings here with this point. More specifically, when discussing inferred audience demographics, we emphasize that non-voters and voters from many marginalized groups are not represented, or are significantly underrepresented, in this sample.

\subsection{Characterizing Longitudinal Publics (RQ1)}
We address RQ1 using qualitative analysis, determining 1) labels for each of the clusters identified (``labels'') and 2) a higher level typology of the labels themselves (``types''). To determine both labels and types, we conducted two rounds of coding, including a first round of open coding by all four authors and a second round of focused coding to finalize labels and types \citep{charmaz_grounded_2008}. Two authors independently coded each cluster based on the top 50 (as determined by factor loadings) creators in that cluster; they then met to determine a final label and consulted a third author as needed to resolve disagreements. In cases where consensus could not be reached, the authors examined more accounts in the cluster. If no consensus could be reached at that point, the cluster was labeled using the phrase NA, or None. Types were determined first by three rounds of open coding in which a single author iteratively refined the coding scheme, then finalized with a round of focused coding with two additional authors. This qualitative process resulted in 150 labeled clusters and associated types. We select a cutoff point --- at most 3,500 creators per cluster --- such that our overall set of creators includes greater than 85\% of the political accounts from \citet{wojcieszak_most_2022} and the elite accounts from \citet{mukerjee_political_2022}. See Appendix~\ref{app:compare} for additional details. The 516,375 accounts that this method captures make up over 41\% of the average panel member's total relationships --- a substantial proportion of the following relationships on the platform for American Twitter/X users (see Figure~\ref{fig:infl_follows} in the appendix). 

To ensure the robustness of our findings we reran the PCA on the follower network of the Twitter Panel half a year and a year after our original data collection in September 2020. Again the clusters were first coded by two researchers independently who then met to discuss the final labels. Finally, they consulted the list of labels from the September 2020 clusters to remove any semantic inconsistencies in cluster labeling. In April 2021 146 out of 150 (97.3\%) identified longitudinal publics showed at least some overlap in the top 50 creator accounts with those found in September 2020. 136 out of them (90.7\%) even showed an overlap of minimum 40 top 50 creator accounts ($\geq$ 80\%). In September 2021 still 145 (96.7\%) clusters showed some overlap with the clusters in September 2020, 131 (87.3\%) contained minimum 40 identical top 50 creator accounts. Our analysis however is based on the follower data collection in September 2020 because this point in time aligns best with the collection of Tweets. In line with findings on the long term stability of following relationships by \citet{zhang_social_2021} we find that longitudinal publics are reasonably stable over time. We provide a more detailed analysis of the overlap in the second level types in Figure~\ref{fig:clus_robustime} in the appendix.

\subsection{Discursive Overlap with Issue Publics: Electoral Politics and BLM (RQ2)}

Our second research question centers around explaining heterogeneity in the engagement of longitudinal publics in 2020 with two specific issue publics---electoral politics and racial justice---and how specific events temporarily shaped those engagements. We cover how we identified engagement with these issue publics and discuss our statistical analyses.

\subsubsection{Identifying Content from Electoral Politics and Racial Justice Issue Publics}

As in \citet{shamir_who_2023}, we use the Twitter/X Decahose, a 10\% random sample of Twitter/X, to discover content produced by the longitudinal publics under study. We specifically look at data from May 7, 2020 (the first day we have Decahose data available) until December 31, 2020. This yields seven months of tweets and encompasses the summer of 2020, in which major protests for and against the Black Lives Matter movement occurred across the country, as well as the fall containing the 2020 U.S. general election. 

We analyze these two distinct issue publics, each of which captures different avenues for civic engagement, using pre-validated keyword lists drawn from prior studies of Twitter/X in 2020. First, we use the Black Lives Matter keyword list from \citet{shugars_pandemics_2021} to quantify when, and how much, creators were discussing the Black Lives Matter movement. \citet{shugars_pandemics_2021} found that accuracy, precision, and recall were above 88\% on a manually coded sample of tweets. Second, we combine the keyword lists used to detect speech related to electoral politics in \citet{gitomer_speech_2023} and \citet{mukerjee_political_2022}. The keywords used by \citet{gitomer_speech_2023} had a precision of above 90\% with recall above 75\% (for an F1 score of over 0.8); the \citet{mukerjee_political_2022} list was not formally evaluated but was used in several prior works. We note that, out of the 345 BLM and 333 electoral keywords, there are only 5 overlapping terms that match exactly: these keyword lists are likely capturing distinct kinds of conversations. 
 
As compared to using a classifier (as in \cite{shamir_who_2023}), use of keyword lists likely underestimates the actual volume of political speech. Ultimately, these lists are both highly effective; are easily connected to prior measurements; and serve as a useful baseline for whether, and when, political discussion is occurring. These keyword lists capture, respectively, potential exposure to discussion of a specific marginalized political movement and generalized discussion of U.S. electoral politics. We are able to measure how many tweets per day by creators matched at least one keyword in each list and compare them to creators' overall tweet volume. We note that we are only measuring the volume of creators' original tweets in the present work, not retweets, quote tweets, replies, and likes; original tweets represent only one mode of engagement on Twitter. This means that we are not measuring behaviors like white allies' amplification of Black voices in the Black Lives Matter movement \cite{clark_white_2019}; users reframing conversations using replies \cite{zade2024reply}; or quote tweeters increasing the reach of political discourse \cite{garimella2016quote}. While we focus on original content generated by a longitudinal public's creators in the present work, amplification and reframing are also arguably forms of communicative output in a longitudinal public.

\subsubsection{Framework for Analyzing Discursive Overlap Between Issue Publics and Longitudinal Publics}

As noted above, we leverage the framework from  \citet{beers_measuring_2025} to measure communicative output of longitudinal publics in a way that considers both creator and audience factors. We focus only on the product term $N_c *A_c$ from Equation~\ref{eq:beers}, or $O$ here, representing content creators' outputs. We then parameterize the total output for each content creator $c$ of a particular public $x$ conditional on 1) a specific issue public $i$, and 2) a specific timepoint $t$. That is, we aim to study different aggregations of $O_{cxit}$, the output of a content creator $c$ in longitudinal public $x$ that is relevant to a specific issue public $i$ at time $t$. We refer broadly to this quantity as the \emph{discursive overlap} between content creators' personal publics within a longitudinal public and a particular issue public at a given time. We use the notation $\sum_{c}$ to represent a summation across one or more facets of the relevant index; for example, we can define $O_{\sum_{ci}xt}$ as the output of \emph{all} content creators within longitudinal public $x$ \emph{all} issues (i.e. the total output) at time $t$.

In analyses for RQ2, we also removed nine longitudinal publics from analysis. Three of these publics had predominantly non-English tweets and thus had language that was not captured by our wordlists for the two issue publics. Six other longitudinal publics were removed because we could not establish the basis of the public on which this cluster of accounts was identified using our methods. Table~\ref{tab:full_categories} in the appendix lists the excluded publics.

\subsubsection{Explaining Heterogeneity over Time (RQ2a)}

For RQ2a, our estimand is the quantity $\frac{O_{\sum_c xit}}{O_{\sum_{ci} xt}}$, the proportion of a longitudinal public $x$'s output at time $t$ related to issue public $i$, where $i$ is either electoral politics or the Black Lives Matter movement. More specifically, we first identify two events, one that we expect will impact longitudinal publics' discursive overlap with the BLM issue public, the murder of George Floyd, and a second event that we expect will impact longitudinal publics' overlap with the Electoral issue public, the 2020 U.S. general election. 

We intentionally selected these two cases for their contrasting predictability/periodicity; further, we note that we do not intend to build a generalizable theory of issue publics' overlap with longitudinal publics from a sample of two events. Sampling cases based on diversity is a reasonable strategy for exploring variation in a causal path \cite{seawright2008case} --- in this case, we compare the overlap dynamics from a predictable, anticipated event (the 2020 U.S. election) to those of George Floyd's unanticipated murder and the widespread protests that ensued, but we do not extrapolate this comparison beyond these two events. 

We construct a window around each event, spanning two weeks prior to the event, the week of the event, and six weeks after the event, and calculate $\frac{O_{\sum_c xit}}{O_{\sum_{ci} xt}}$ for each combination of issue, event, and week. In total, then, we have 36 observations for each longitudinal public (9 weeks x 2 issue publics x 2 events). We then estimate factors that are associated with heterogeneity in $\frac{O_{\sum_c xit}}{O_{\sum_{ci} xt}}$, using a series of linear mixed effects models that account for successively more potential sources of heterogeneity, all with a random effect for each longitudinal public. 

\subsubsection{Explaining Heterogeneity Across Publics (RQ2b)}

We also attempt to explain hetereogeneity in overall engagement with issue publics. In the quantitative language introduced above, our estimand is the quantity $\frac{O_{\sum_{ct} xi}}{O_{\sum{cit} x}}$, the proportion of a longitudinal public $x$'s output over all time periods related to issue public $i$. We have only two observations for each longitudinal public (discursive overlap with the 2 issue publics). To better understand these patterns, we look both descriptively, exploring in detail specific longitudinal publics that have high (or low) engagement with each issue public, as well as quantitatively, focusing on associations between a longitudinal public's type and audience make-up and their discursive overlap with each public. With respect to audience characteristics, we focus on a minimal set of representative demographic variables that can be reliably measured using our data: fraction male, average age of audience, and proportion of audience that is a Republican vs. a Democrat.

Table~\ref{tab:regression_rq2b} presents results from a linear regression (with clustered standard errors on longitudinal public) for a base model that explains $\frac{O_{\sum_{ct} xi}}{O_{\sum_{cit} x}}$ using only an interaction term between longitudinal public type and issue public type, and a full model that incorporates interactions between all three audience demographic variables and issue public. This full model explains the data significantly better than the base model ($p < 0.0001$) and is therefore used to outline results. Further details on these models can be found in Appendix~\ref{sec:2bmethods}.

\subsubsection{Explaining Heterogeneity \emph{Within} Publics (RQ2c)}

Finally, we look quantitatively at heterogeneity \emph{within} publics, focusing on the estimand $\frac{O_{\sum_t cxi}}{O_{\sum_{it} cx}}$, the proportion of \emph{each content creator's} discursive overlaps with each issue public. We restrict ourselves to content creators for whom we have at least 5 tweets (on any topic) in the Decahose during our time window and who have at least 1000 followers; this ensures we are focusing on accounts that meet established minimum thresholds for influential users \citep{starbird_influence_2023}.

Both \citet{beers_measuring_2025} and \citet{zhang_social_2021} explore heterogeneity in expression \emph{across} publics but not \emph{within} them. Within-public variation is theoretically interesting for three reasons. First, disaggregation allows us to characterize the extent to which a longitudinal public's constituent personal publics reflect the longitudinal public's aggregate behavior. Substantial internal variation within longitudinal publics implies that in Equation \ref{eq:beers}, it is necessary to represent both $A_c$ and $A_a$ (creator and audience activity, respectively) as distributions rather than as fixed quantities. More thoroughly explaining heterogeneity within these distributions may require adding more complexity to the model to explain this heterogeneity. 

Second, clues as to which accounts within longitudinal publics are more likely to engage with political issue publics may help us to better understand who brings politics into apolitical spaces. Whether one sees this as normatively good \citep{jackson_ferguson_2016} or bad \citep{talisse_overdoing_2019}, understanding these pathways is important. Users who act as bridges between partisan factions \citep{garimella2018political} or gatekeepers to extremist communities \citep{russo_spillover_2023} intervene in structures of attention and pathways to political expression. Learning more about who drives political spillover, and how, helps us understand pathways by which attention and expression can be shaped by human agency. 

Finally, the extent to which factors associated with personal publics explain (or don't) discursive overlap with issue publics above and beyond longitudinal public membership provides insight into how much explanatory power our theoretical construct has in the context of content creator engagement with political issue publics. To this end, we turn to a fixed effects modelling strategy, a common approach in econometrics when we wish to explain heterogeneity in multi-level modeling settings where we aim, in particular, to partial out variation at higher levels with explicit, parameterized, entity-level intercepts. Further information on this model can be found in Appendix~\ref{sec:2cmethods}.

\section{Results}\label{results}

\subsection{RQ1: What Longitudinal Publics Existed on Twitter in 2020?}

\begin{table}[ht]
\centering
\footnotesize
\renewcommand{\arraystretch}{1.1}
\begin{tabular}{p{3cm} p{7.6cm}}
\toprule
\textbf{All 2nd Level Labels} & \textbf{Longitudinal Publics} \\
\midrule

\multicolumn{2}{l}{\textit{National Politics}} \\
Political & Mainstream Left-Leaning, Progressive Liberals, Political Republicans \\

\addlinespace
\multicolumn{2}{l}{\textit{Local News/Politics}} \\
Local \& Political & Chicago Politics, Minnesota Government \& Adjacent, Philadelphia News \\

\addlinespace
\multicolumn{2}{l}{\textit{Others}} \\
Business & Venture Capitalists/strategists, Crypto, Real Estate \\
Sports & Ultimate Fighting Championship, Cycling, Pro Poker,  Women’s Basketball, Sprint Car \\
Education (\& Business) & Administrative K--12 Education, Innovators in Education, Math Ed \\
Religious & Christian Leaders, Catholics, White Christian Women Authors \\
Lifestyle & Food, Weddings, Wine, Interior Design/Furnishings \\
Celeb/Influencer & Late Millennial Celebs, Generic Influencers, Black Influencers \\
Entertainment & Outlander Series Enthusiast, Children’s Authors/Illustrators \\
No Clear Pattern & NA / Bot, NA/Non-US, None \\

\bottomrule
\end{tabular}
\caption{Longitudinal Public types (left column) and examples of these types (right column)}
\label{tab:short_categories}
\end{table}

\begin{figure}[!t]
    \centering
    \includegraphics[width=\linewidth]{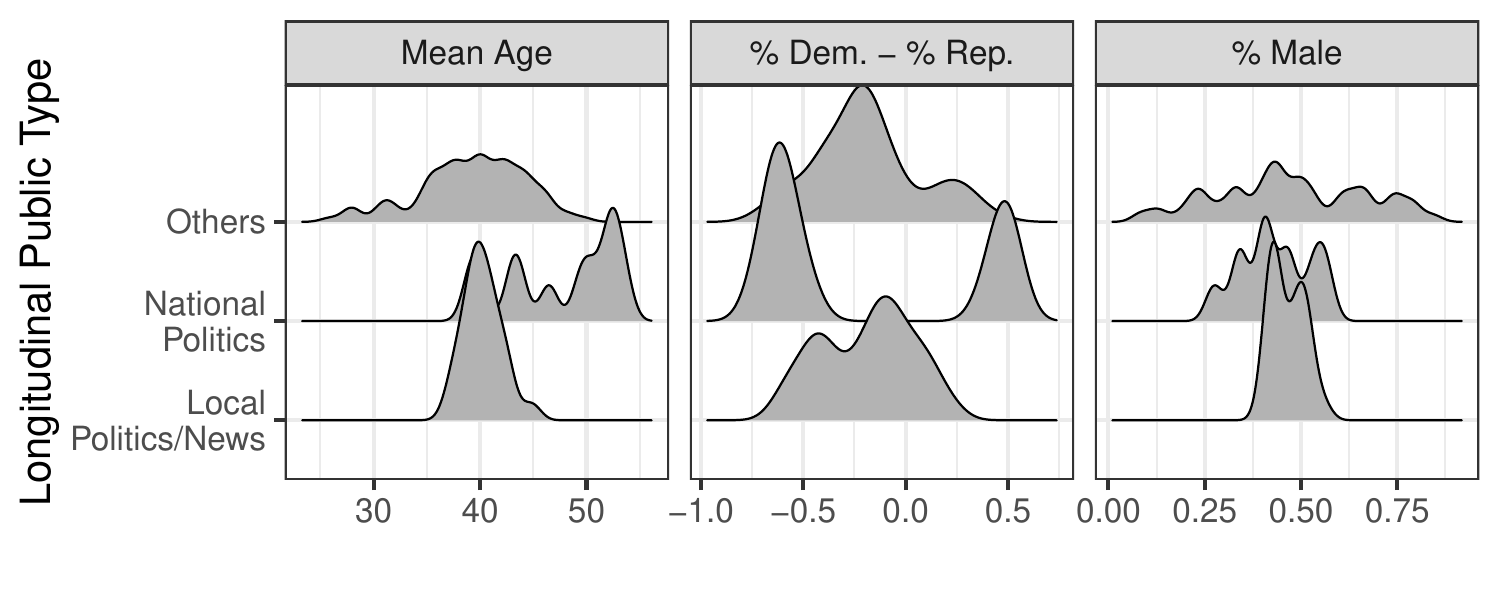}
    \caption{Density plots for the three different audience demographics metrics we analyze (different subplots) for the three overarching longitudinal public types we look at (y-axis) }
    \label{fig:aud_demog}
\end{figure}

Longitudinal publics emerge in our analysis that represent a range of intersecting interests, places, and identities. In the context of studying overlaps with issue publics we find it useful to categorize them into three high-level categories: National Politics, Local News \& Politics, and all other categories, which comprise the online third spaces that are of particular interest here. Table~\ref{tab:short_categories} displays labels for a subset of longitudinal publics (right column) along with second-level types that we assigned to them (left column) and the highest-level categorization most relevant to our analysis of discursive overlap (subheaders within the table). Table~\ref{tab:full_categories} in the appendix provides the full set of longitudinal publics and public types we identified.

We identify eleven longitudinal publics where content creators primarily identify with national-level political discourse, twenty-four where the focus is local but still primarily centered around hard news and/or politics, and 115 that are not explicitly political. As such, while longitudinal publics focusing on news and politics serve as a useful comparison point for our case study, the majority of longitudinal publics we identified can be understood as online third spaces grounded in distinct cultural, geographical, and/or identity-based foci. 

With respect to audience makeup, we find that these high-level categories vary in important ways that reinforce the potential importance of online third spaces in promoting attention to cross-partisan political discourse. Figure~\ref{fig:aud_demog} shows that national politics-focused longitudinal publics have a bimodal political distribution and skew older; local news and politics show relative parity along all demographics; and the ``other'' category shows significant spread across audience characteristics. This provides concrete and large-scale empirical evidence for the claim made by \citet{talk2015third} that online third spaces may offer more heterogeneity in terms of political opinion, as well as increased opportunities for contact with weak ties, compared to political spaces. 

Moving beyond measures of audience demographics, we find longitudinal publics centered on culture, place, and identity. With respect to culture, across business, sports, and entertainment, publics emerge around content creators who share a central space in the cultural landscape. These cultural issue-centered longitudinal publics sometimes emerged around unanticipated (at least by the authors of this paper) foci. For example, we identify a cluster where influencers focused almost entirely on golf turf, including several golf turf experts and accounts promoting degrees in golf turf. Such publics would be unlikely to be the focus of content-driven analysis but nonetheless demonstrate the variety of structurally identifiable third spaces that can emerge from stable relations of audiences and their creators.

Place-oriented longitudinal publics often centered around local news and politics: 24 of the 35 publics that we identified as Local were also classified as Politics and/or News. But within these clusters, politics and news were interleaved with the non-political, either with content creators that identified more broadly around news (e.g. local news stations) or with a small number of creators centered on apolitical topics centered within predominantly news and politics (e.g. accounts for sports talk radio). These clusters also varied in both the attention devoted to specific topics (e.g. politics versus sports) and their geographic spread. With respect to geographic spread, several clusters grouped together influential accounts at the state level (e.g. for Hawaii, Kentucky, and Minnesota), while others centered on specific cities, both large (e.g. Chicago, Philadelphia) and small (e.g. Buffalo, Cincinnati).  These spaces demonstrate that the ``political-ness' of longitudinal publics is best understood on a continuum, rather than bifurcated between the non-political third spaces and overtly political arenas of discourse. 

Finally, we identify longitudinal publics that represent specific social identities \citep{burke_self_2003}, thus suggesting that longitudinal publics can sometimes take the form of more community-driven overlapping personal publics around both shared relations and a shared identity. For example, we identify a cluster consisting almost entirely of managers of Home Depot stores across the country. Here, the identity of managing a Home Depot, rather than a particular interest in doing so, best characterizes this behavior. Similarly, we identify clusters where personal publics are centered around leaders in Christianity, and another for Catholicism more specifically, as well as a longitudinal public consisting of accounts presenting, at least, as emergency medicine doctors. These all are defined best by a clear social and/or collective identity. Finally, our qualitative work suggested that several of the clusters we uncovered represented likely to function as networked counterpublics for marginalized identities. This included clusters centered on environmental activism and a broader social justice-oriented cluster with a particular (though not exclusive) focus on race. 

Of course, identity, place, and interest are inter-related. For example, individuals' interests can define elements of their identity, as in the context of gaming \citep{gray_intersectional_2020}, and sports are an example of how place can shape which interest clusters are attended to. Someone who is interested in sports generally in Buffalo versus Miami will attend to different sports-related clusters, because these cities have professional teams in different leagues. Perhaps more interesting empirically are clusters that intersect place, interest, and/or identity. Examples of these intersectional clusters include one of White Christian Women Authors, intersecting interests and both gender and religious identities, and one of Mormon leaders, Utah media, and Brigham Young University sports that intersects place, religious identity, and an interest in sports. 

\subsection{RQ2a: Explaining Heterogeneity in Discursive Overlap over Time.}

\begin{figure}[!t]
    \centering
    \includegraphics[width=\linewidth]{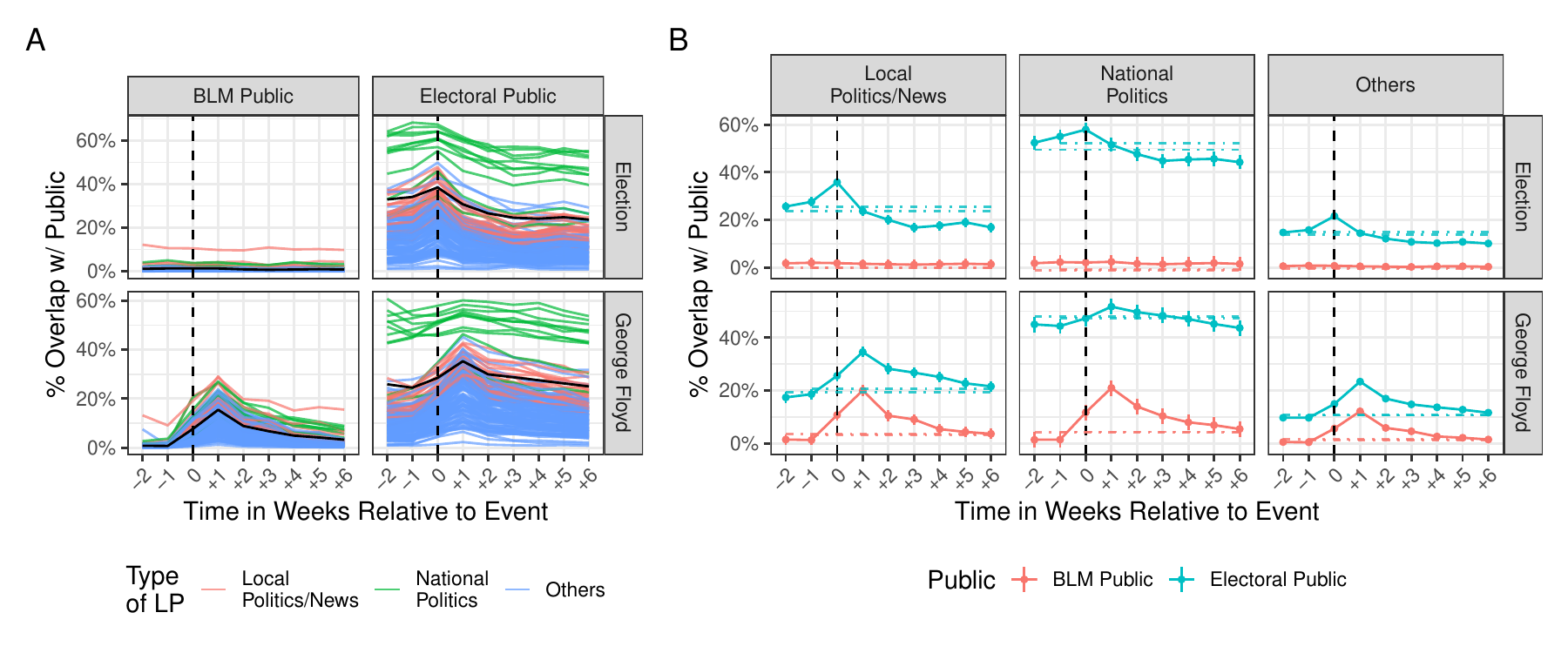}
    \caption{A) Discursive overlap (y-axis) over time (x-axis) for each longitudinal public (separate colored lines, colored by high-level type) for each combination of issue public and event (different subplots). The black line represents the mean over all longitudinal publics for each week.\\
    B) Marginal effects with 95\% confidence intervals from our full regression model for discursive overlap (y-axis) over time (x-axis) for the two different events we analyze (different rows) for each type of longitudinal public (different columns) for two different issue publics (color). The black vertical dotted line is a visual cue for the week of the event, and the two horizontal colored lines are the higher (lower) end of the 95\% confidence intervals for the estimated proportion of discursive overlap for the two week before the event for George Floyd (the election). These are added to show differences relative to this level in the post-event weeks. }
    \label{fig:rq2_time}
\end{figure}

The two events we studied drive similar overlaps between issue and longitudinal publics in their immediate aftermath, but differences exist in how long those overlaps last and how much other discourse is displaced by issue public-centered discourse within longitudinal publics. Figure~\ref{fig:rq2_time}A shows discursive overlap across issue publics and events for each longitudinal public separately, providing intuition that guides our statistical analysis. Our statistical results are presented in Figure~\ref{fig:rq2_time}B in the form of marginal effects from our regression model. As expected, the murder of George Floyd drove a significant increase in overlap, on average, between longitudinal publics and the BLM issue public. Similarly, as expected, the 2020 General Election drove, on average, a significant increase in overlap between the Electoral public and the longitudinal publics we analyze. 

Figures~\ref{fig:rq2_time}A and B show, however, that in contrast to the 2020 U.S. General Election, discursive overlap with both the Electoral issue public and the BLM issue public remained significantly higher than baseline several weeks after Floyd's murder across all types of longitudinal publics. As we discuss further below, this emphasizes the need to explore the ways in which events beyond elections shape the electoral publics conceived of by \citet{beers_measuring_2025}.

This general temporal pattern varies substantially across even the high-level types of longitudinal publics considered here. Indeed, we find that the murder of George Floyd led to a much more sustained discursive overlap between the Electoral issue public and both local news/politics and other longitudinal publics compared to national politics-centered longitudinal publics. The mean estimated level of discursive overlap between the national politics-centered publics and the electoral issue public fell within the 95\% confidence interval of pre-George Floyd levels within two weeks after his murder, whereas this discursive overlap remained elevated above this level for five (local politics/news) to six (others) weeks after the event in the other cases.  While such hetereogeneity is expected \cite{zhang_social_2021}, we now turn to explore the factors associated with this heterogeneity in discursive overlap with these two issue publics occurs across longitudinal publics.

\subsection{RQ2b: Explaining Heterogeneity in Discursive Overlap Across Longitudinal Publics}

\begin{figure}[!t]
    \centering
    \includegraphics[width=\linewidth]{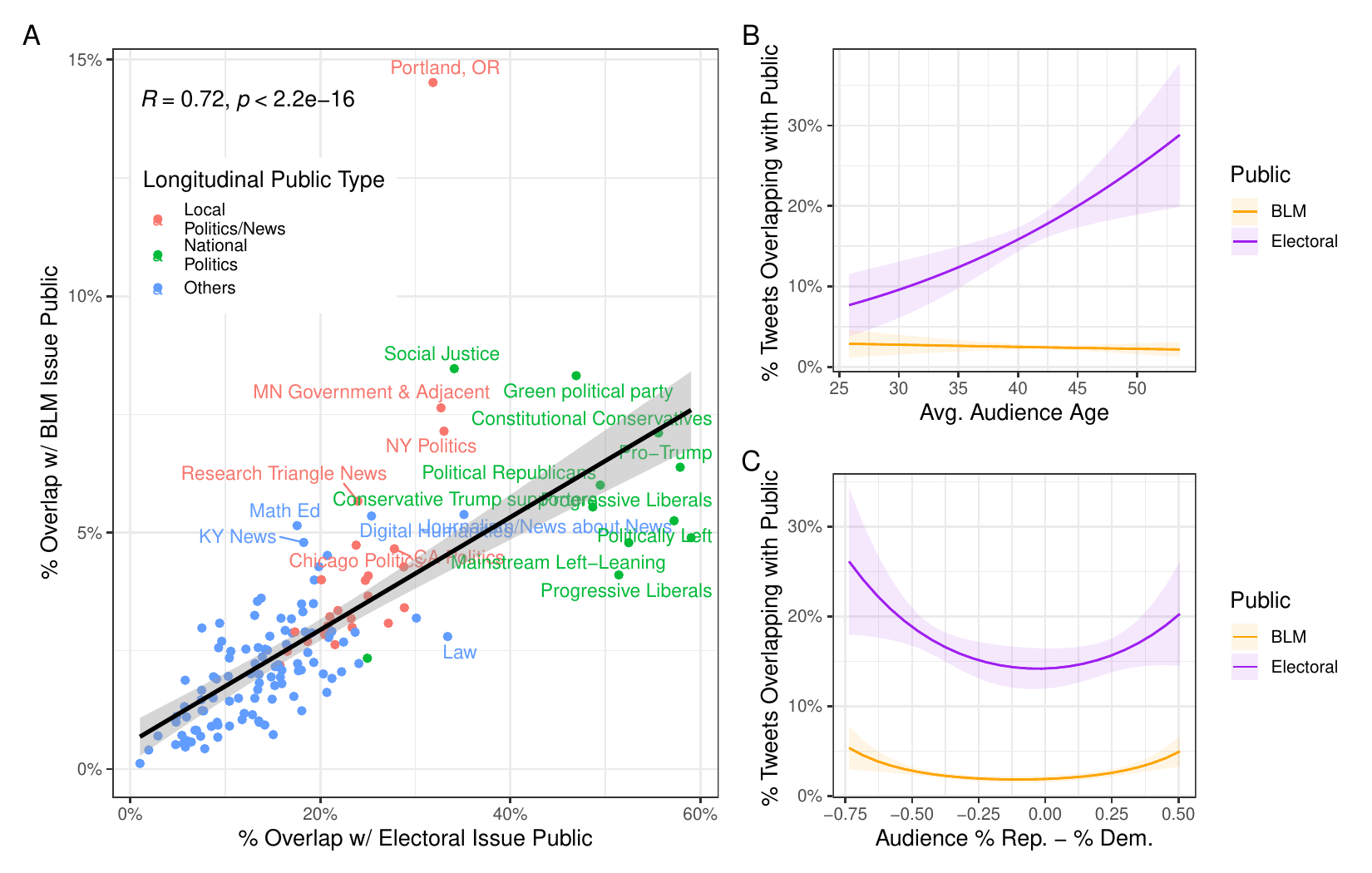}
    \caption{A) The average proportion of discourse overlap between each longitudinal public (a dot in the figure) and the Electoral (x-axis) and BLM (y-axis) issue publics. Point are labeled while avoiding overcrowding of labels, and colored by their high-level longitudinal public type.\\
    B and C) Marginal effects with 95\% confidence intervals for discourse overlap (y-axis) over time (x-axis) for the three audience demographics we consider (different plot rows) for the three different types of publics we study (plot columns) for two different issue publics (color). Marginal effects are computed from our full statistical model, detailed in the text.}
    \label{fig:rq2_overview}
\end{figure}

Our high-level typology explains most of the variation \emph{across} longitudinal publics, but audience-level factors do have additional explanatory power: model fit improves significantly ($\chi^2(1) = 587934, p <.0001$) with the additional variables, as do model AIC and BIC. Figure~\ref{fig:rq2_overview}A shows the proportion of each public's output that overlapped with each issue public. Several points are immediately apparent.  First, publics we categorize as National Political news have the highest discursive overlap with electoral issue publics, followed by local politics and news, and then all other publics. However, this is not true for the BLM issue public, where engagement was more heterogeneous across public types. Moreover, considerable heterogeneity existed within each type. We also observe clear outliers. Most notably, local news and politics relevant to Portland overlaps considerably more with the BLM public than any other longitudinal public, but other outliers also exist. 

The audience makeup of longitudinal publics explains some of this heterogeneity within public type, especially for the Electoral issue public.  Table~\ref{tab:regression_rq2c} shows full regression results; here we consider only those that are both statistically and practically significant. Figure~\ref{fig:rq2_overview}B shows marginal effects from our regression model for the two audience-level factors that are both statistically and practically (in terms of average effect) significant: average audience age and the proportion of the audience inferred to be Democrats vs. Republicans. We note here that due to the way the panel audience members was constructed, non-voters and voters from many marginalized groups are not represented, or are significantly underrepresented, in this sample. Figure~\ref{fig:rq2_overview}B shows that adjusting for longitudinal publics' other audience characteristics, overlap with electoral issue publics increased, on average, from around 8\% to around 29\% moving from the lowest to the highest observed average audience age in our dataset. However, there was almost no association between mean age of longitudinal public audience and overlap with the BLM issue public. We see limited evidence of a statistically or practically significant association between partisanship of a longitudinal public's audience and engagement with the BLM issue public, but a more extreme partisan makeup on either side of the political spectrum was associated with increased overlap with the Electoral issue public. This is, again, suggested by the marginal effects plots in Figure~\ref{fig:rq2_overview}C.

\subsection{RQ2c: Explaining Heterogeneity Within Publics in Engagement}

\begin{figure}[!t]
    \centering
    \includegraphics[width=\linewidth]{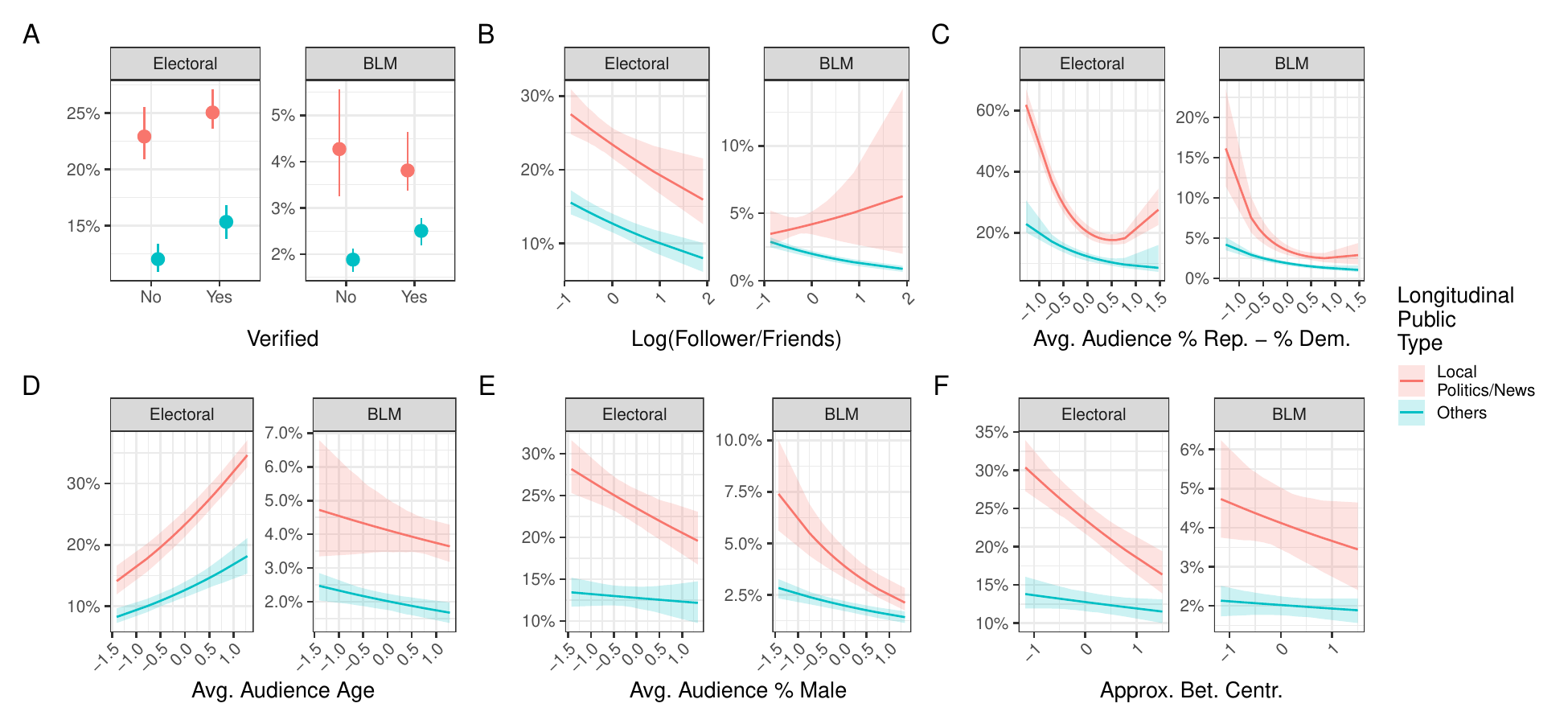}
    \caption{A) Marginal effects estimate for whether or not an account was verified (x-axis) on discursive overlap (y-axis).\\
    B-E) Marginal effects estimates for different measures of demographics for the personal publics of different audience creators (labels on the x-axis).\\
    F) Marginal effects estimates for approximate current flow betweenness centrality (x-axis) on discursive overlap.\\
    For all plots, effects are split between the different issue publics (separate subplots) and across longitudinal public types (colors); we do not show results for the National Politics public type due to small N. Marginal effects are shown over averages of all other variables, where continuous variables are partitioned into 20 intervals for the computation. Error bars are computed using a clustered bootstrap proceedure with N=1000 bootstrap replications.
    \\
    For plots B-F), x-axis represent scaled and mean-centered quantities (see Methods section), and we only show x-axis representing values for the middle 95\% of all data points.
    }
    \label{fig:rq2c}
\end{figure}

Knowing which longitudinal public a creator belongs to explains most variation in engagement with issue publics; a model containing only these factors has an adjusted $R^2$ of 0.89. However, factors associated with the creator themselves, and their own individual personal public, can help explain within-public variation. Adding these coefficients shows a limited improvement in adjusted $R^2$ (to 0.90), but improves model BIC and root mean squared error (RMSE), and is able to explain 11.2\% of the variation \emph{within} publics. Substantively, they also help us to consider how individual creators may make decisions within the context of their own personal public. 

Figure~\ref{fig:rq2c} displays marginal effects for the regression model used to study associations between attributes of creators and their personal publics and engagement with issue publics; full details can be found in Appendix~\ref{sec:2cmethods}. Marginal effects for publics focused on National Politics provided limited practical significance, and so are not displayed.

With respect to creator attributes indicated by Figures \ref{fig:rq2c}A and \ref{fig:rq2c}B, differences in verification status and audience size between the creators that comprise a longitudinal public were associated with differences in the extent to which a creator's personal public overlapped with an issue public. In the ``Others'' category, for example, less popular verified accounts demonstrated more discursive overlap with both electoral politics and BLM; in local politics/news longitudinal publics, this was also true for electoral politics, but more popular, unverified accounts had more discursive overlap with the BLM issue public. The effects of verification and audience size were limited for longitudinal publics dedicated to national politics. 

With respect to attributes of the creators' personal publics, as Figures \ref{fig:rq2c}C-E indicate, the effects of audience demographic attributes on discursive overlap between personal and issue publics are also heterogeneous. We note here that due to the way the panel audience members was constructed, non-voters and voters from many marginalized groups are not represented, or are significantly underrepresented, in this sample. Creators with left-leaning audiences experienced more discursive overlap overall, but this effect is ameliorated in the case of an extremely right-leaning audience for local news/politics. Older audiences were associated with more overlap with electoral issue publics, while the opposite was true for BLM. Within local news/politics longitudinal publics and other longitudinal publics, creators whose audiences that skewed more male had less overlap with issue publics; however, more male audiences were associated with a slight increase in overlap with electoral publics for national politics longitudinal publics. And, as we see in Figure \ref{fig:rq2c}F, creators whose accounts were more central within a longitudinal public demonstrated less overlap with issue publics.

\section{Discussion \& Conclusion}\label{sec12}

\subsection{Theoretical Contributions}
The primary theoretical contribution of the present work is the longitudinal public, which synthesizes prior contributions by \citet{brunsPublicSphereNetwork2023}, \citet{beers_measuring_2025}, and especially \citet{zhang_social_2021}, to ask different questions than each would enable alone. In outlining longitudinal publics and characterizing what binds them together, we are able to discover publics on social media that differ substantially from the flocks discovered by \citet{zhang_social_2021}. A prerequisite for flock membership is being followed by influential actors \citep{zhang_social_2021}, but longitudinal public membership simply requires that creators command the attention of a set of users that is representative of a population that is meaningfully defined offline and have posted at least 5 tweets in the period we studied. Rather than providing information about variations in elite opinion, then, as we would see with the flock construct, the longitudinal publics we discover are indicative of the ways that identity, place, and culture interact in U.S. Twitter users' social media landscapes. In the same vein, the flock construct is not designed to discover third spaces, where shared identity might be based on a shared artistic practice, career path, or sports fandom, rather than political partisan divisions. Allowing for third spaces allows us to more effectively integrate our ideas into the overarching model from \citet{brunsPublicSphereNetwork2023} by introducing the concept of \emph{discursive overlap}, where an issue public bleeds into a public that is not necessarily explicitly political. Discursive overlap offers a way to think about the bridges, both theoretically and concretely, between publics defined via content, like issue publics, and publics that are defined relationally, like longitudinal publics. We consider political output by creators in spaces that are not necessarily political through the lens of discursive overlap: how does the political become personal in a distinct space formed around a specific focal point --- in a world made and remade discursively \citep{warnerPublicsCounterpublics2021}?

We build on the model put forth by \citet{beers_measuring_2025}, as well as her call for future work on the ``how'' and ``why'' of the model's change metrics, to understand what influences creators' political output in a public (sphere). Based on our findings, we believe that it is necessary to conceive of the quantity $A_c$ (creators' activity level) as a distribution, and this is likely true of other quantities in the model as well, such as $N_a$ or $A_a$ (audience size and sharing rates). The differences we find across longitudinal publics' discursive overlap with issue publics explains a great deal of the variation we observe, but substantial unexplained heterogeneity still exists \emph{within} longitudinal publics' creator populations. Our findings suggest that future work studying the extent to which audience behavior within a longitudinal, issue, or election public helps explain variations in a public's overall discursive output may provide further corroboration and contextualization for Beers' model. While this work focuses primarily on creator behavior and audience \emph{characteristics}, understanding what explains audience behavior within a public (sphere) may help us understand the extent to which creators' decisions with respect to content creation might shape audience behavior and, therefore, public opinion more broadly.

\subsection{Empirical Contributions}
By discovering the longitudinal publics presented here, we are able to describe the kinds of discursive spaces that U.S. voters on Twitter have long-standing relational connections to, regardless of whether those spaces engage regularly with politics, and reason about the points at which issue or election publics might bleed into these spaces. We corroborate findings by \citet{beers_measuring_2025} that temporally limited electoral publics do emerge consistently around elections. However, our analysis of longitudinal publics shows that electoral/issue publics are potentially reshaped by other political events and worth studying at these times as well. Events can differentially reshape longitudinal public discourse over time, sometimes displacing other discourse (as in the case of George Floyd's murder), and other times yielding room for new discourse to emerge (as with the end of the 2020 election cycle). \citet{brunsPublicSphereNetwork2023} emphasizes studying points where personal and issue publics overlap; we find that longitudinal publics, as aggregations of personal publics, can and do experience discursive overlap with electoral/issue publics in ways that are shaped by offline events. Election and issue publics are time-bounded, either intermittently or confined to a single event \citep{beers_measuring_2025, brunsPublicSphereNetwork2023}, and observing how they bleed into durable longitudinal publics that coalesced because of common locality, hobbies, identities, or interests helps us understand how political discourse might become personal. 

We find that, at the level of an entire longitudinal public, three clear trends arise. First, we find that online third spaces --- those not \emph{a priori} dedicated to political discourse --- have politically diverse audiences, creating opportunities for cross-partisan contact as was hypothesized by \citet{talk2015third}. Second, overlap between longitudinal and issue publics is, as expected, heavily driven by relevant events. However, we see that 1) the aftermath of these events produces different patterns in sustained engagement, and 2) that events not inherently related to a public---in our case, George Floyd's murder and the electoral issue public---can have significant, and even greater, impacts on the spread of an issue than the directly relevant event itself. We do not believe that the lasting surge in overlap with the electoral issue public indicates confounded outcome variables, particularly in light of very low keyword overlap between the two issue publics, so much as it captures the fact that, in line with findings by \citet{caren2025black} and \citet{klein2025weather}, racial justice had become a campaign issue and entered mainstream political discourse. Finally, audience demographic makeup predicts overlap with the electoral issue public, but this predictive power is largely limited to average age and partisan makeup from the factors we considered, and it has little predictive power for overlap with the BLM issue public. Given this, viewing publics like the flock or the longitudinal public as heterogeneous makes sense: audiences and creators within a given public do not behave monolithically. When discussing inferred audience demographics, we are careful to point out that due to the way the panel of U.S. voters who are also Twitter/X users was constructed, non-voters and voters from many marginalized groups are not represented, or are significantly underrepresented, in this sample.

Within longitudinal publics, we are able to partially explain heterogeneity in individual creators' discursive overlap with election and issue publics. In longitudinal publics not dedicated to local and national politics, less popular verified accounts exhibited more discursive overlap. Perhaps this indicates that established creators with niche audiences had less to lose when bringing politics into seemingly apolitical spaces. It may be that a smaller audience dedicated to a niche creator is more easily imagined by that creator as receptive to political discourse \citep{litt_imagined_2016, su2022social} While our findings also indicate a greater inclination for discursive overlap in apolitical longitudinal publics for creators with left-leaning audiences, we see some increase in  discursive overlap for local longitudinal publics as audiences become very right-leaning as well. Perhaps audiences with less partisan skew belong to local creators who engage with topics like sports or the local food scene, where discursive overlap with issue publics may be less acceptable and more jarring to audience members: negotiating the political in the context of the personal looks different depending on which aspect of the personal is at stake \citep{hartley2006public}. Further, we note that more central creators within the network of a given longitudinal public had less overlap with issue publics, perhaps suggesting some sort of negative feedback loop between prominence in a community and individual political discursive output. This may offer an interesting twist to the literature on discursive gatekeepers \cite{garimella2018political} and bridges \cite{russo_spillover_2023}, if influence in a community and willingness to bring in political discourse are negatively correlated through some causal mechanism. Is this because bringing in politics reduces most users' influence, or is it because influential users are hesitant to discuss politics and potentially risk their status within a particular space? Taken together, our findings indicate that creators' particular relational status with their audiences, coupled with incentive structures that vary between election/issue publics, offer insight into what drives political spillover into apolitical spaces. 

\subsection{Future Work}
The longitudinal public construct has utility beyond Twitter/X and the panel of U.S. voters that was used to define the audience in the present work. In any situation where it is possible to observe the online attention patterns of a representative sample of a well-defined population whose boundaries are meaningful offline as well as online, it is also possible to discover that group's particular landscape of longitudinal publics. Infrastructures and approaches for data donation (see for example \citet{yangCouplingGDPRData2024, yapDigitalDataDonation2025, zannettouAnalyzingUserEngagement2024}) are particularly promising for the discovery of longitudinal publics encountered by specific groups of users, such as at-risk adolescents, disabled adults, or older adults who share misinformation. While the follower relationship is less meaningful on short-form video platforms like TikTok or YouTube shorts, and algorithmic curation dominates, donated watch history data would nonetheless likely provide enough information about an audience's attention patterns to construct coherent longitudinal publics for further observation. 

Network sampling techniques, coupled with knowledge of a population, could also be use to identify meaningful audiences on social media platforms. For example, spikes in Bluesky account creation followed the 2024
restriction and ban of Twitter/X in Brazil
\cite{smithBlueStartLargescale2026}; modern network sampling methods
\citep{cui2022survey, mouw2012network, papagelis2011sampling},
combined with language detection of Brazilian Portuguese posts
\cite{stiilpen2016methodology}, could locate and characterize this
influx---and, because the Bluesky API exposes timestamped follow
events \citep{smithBlueStartLargescale2026}, could support a fully
temporal extension of the longitudinal public construct. Finally, some of our results suggest interesting directions for future qualitative work. For example, we find that the longitudinal public dedicated to local news and politics in Portland, Oregon was an outlier with respect to its discursive overlap with the BLM issue public. An in-depth qualitative study of this overlap could join work by \citet{bickel-knitting-2020} in explaining how shared social context might bring about and sustain surges in particular political discourses in an online space. Overall, we think that the longitudinal public can and will travel beyond Twitter/X and this first use case; researchers who want to study the intersections between the personal and the political, the creator and their audience, or the individual and the collective, would benefit from considering this construct.

\backmatter

\section*{Declarations}

\subsection{Availability of data and materials}
Due to data use agreements (e.g., voter file linkage data), the data used in this study cannot be publicly shared. Aggregated or derived data may be available from the authors upon reasonable request, subject to restrictions
\subsection{Competing Interests}
Not applicable.
\subsection{Funding}
KJ and HS were supported in part by a grant from the National Science Foundation, \#IIS2145051. AHS was supported in part by the National Science Foundation Graduate Research Fellowship Program under Grant No. 1938052. Any opinions, findings, and conclusions or recommendations expressed in this material are those of the authors and do not necessarily reflect the views of the National Science Foundation. JG was supported by the Deutsche Forschungsgemeinschaft (DFG, German Research Foundation) under Project No. 552234710.
\subsection{Authors' contributions}
All authors contributed to the conceptualization and implementation of the methods for the study. AHS and KJ additionally conducted the writing of the paper.
\subsection{Acknowledgements}
Not applicable.

\bigskip
\begin{flushleft}%
Editorial Policies for:

\bigskip\noindent
Springer journals and proceedings: \url{https://www.springer.com/gp/editorial-policies}

\bigskip\noindent
Nature Portfolio journals: \url{https://www.nature.com/nature-research/editorial-policies}

\bigskip\noindent
\textit{Scientific Reports}: \url{https://www.nature.com/srep/journal-policies/editorial-policies}

\bigskip\noindent
BMC journals: \url{https://www.biomedcentral.com/getpublished/editorial-policies}
\end{flushleft}

\begin{appendices}
\section{Comparisons with Prior Work} \label{app:compare}

\begin{figure}[!t]
    \centering
    \includegraphics[width=.97\textwidth]{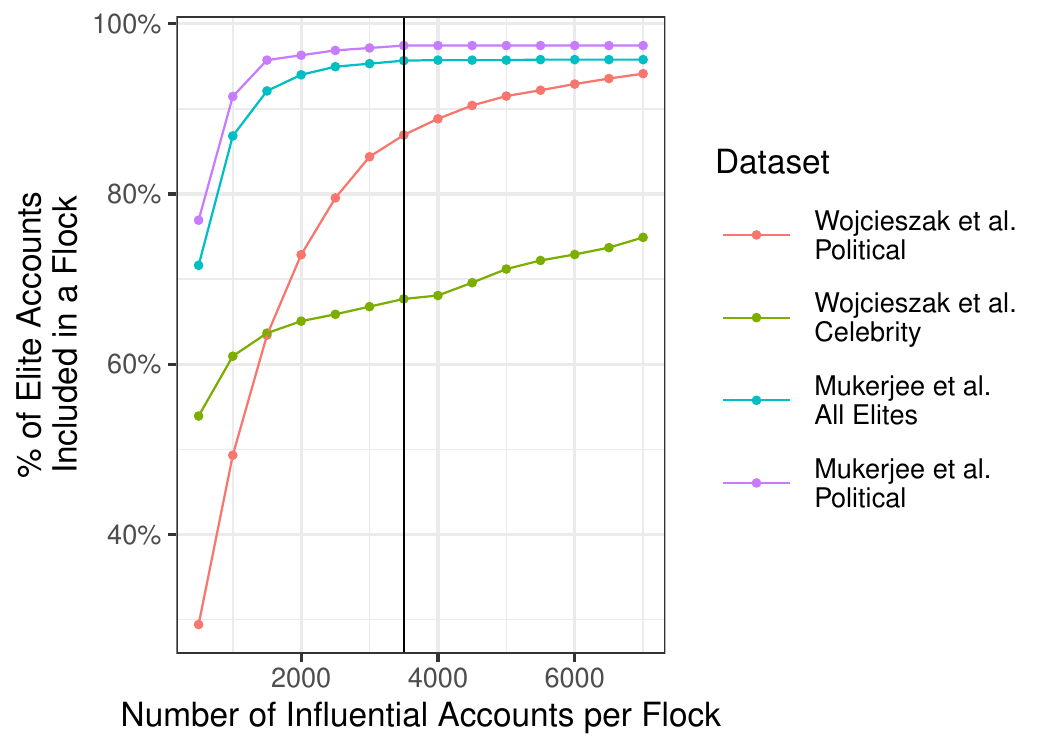}
    \caption{The percentage of accounts (y-axis) in four different lists (the list of 1) celebrities and 2) political elites from \citet{wojcieszak_most_2022}, and the list of 3) political influential users and 4) all elites from \citet{mukerjee_political_2022}) that are represented in a flock when we set the cutoff for flock sizes to the given value on the x-axis. A vertical line is drawn at the cutoff selected for the paper (3500).}
    \label{fig:cutoff}
\end{figure}

Figure~\ref{fig:cutoff} presents our analysis of the correct cutoff point for the number of influential users to include in each flock. We use 3,500 as the cutoff, which gives us 85\% or more of the accounts used in lists from prior work. The only exception is the list of celebrities from \citet{wojcieszak_most_2022}. Much of this list appears to be out of date, even with respect to our analysis in 2020. We therefore exclude it from any analyses in the main paper, and from our cutoff assessment.

\section{Distribution of Follow Relationships to creators Analyzed}

\begin{figure}[!t]
    \centering
    \includegraphics[width=.8\textwidth]{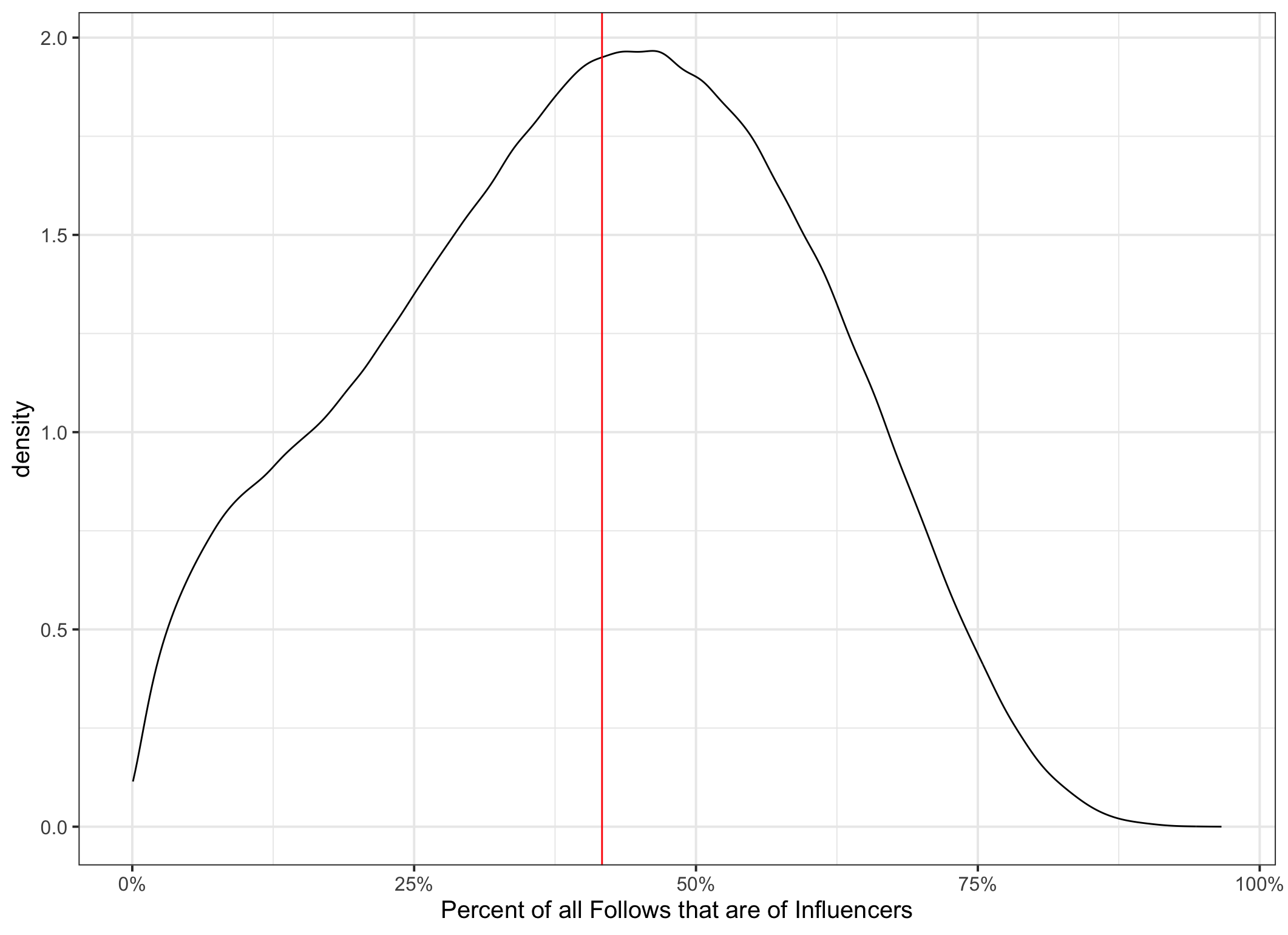}
    \caption{Distribution of the share of panelist follows that are to creators.}
    \label{fig:infl_follows}
\end{figure}

Figure~\ref{fig:infl_follows} displays the distribution across all panel members, where for each panelist we compute the percentage their following relationships that are directed towards one of the 516,375 creators analyzed in this paper. The red line displays the median value (41.9\%), the mean of the distribution is 41.4\%. Over 34\% of panelists (34.3\%) had over half of their following ties directed towards creators analyzed here.  

\section{Full Table of Twitter Longitudinal Publics}

\begin{center}
\centering
\footnotesize
\renewcommand{\arraystretch}{1.}
\begin{longtable}{ p{3cm} p{7.6cm} }
\toprule
\textbf{All 2nd Level Labels} & \textbf{Longitudinal Publics} \\
\midrule

\multicolumn{2}{l}{\textbf{National Politics}} \\
\midrule
Political & Mainstream Left-Leaning, Social Justice, Environmental Activists, Progressive Liberals, Political Republicans, Green Party, Constitutional Conservatives, Politically Left, Pro-Trump, Conservative Trump Supporters \\ 
\addlinespace
\multicolumn{2}{l}{\textbf{Local News/Politics}} \\
\midrule
Local \& Political & Chicago Politics, California Politics, Minnesota Government \& Adjacent, Florida Politics, Massachusetts Gov’t and News, New York Politics \\ 
Local \& News & St. Louis News/Media, Hawaii Media, Philadelphia News, Portland, Oregon, Kansas City News/Media, Oklahoma City News/Media, Cincinnati, Wisconsin Media, Research Triangle News, Detroit \& Michigan News, Denver News/Politics, Indianapolis News/Media, Phoenix, Arizona News \& Gov’t, Texas News, Pittsburgh Media, Memphis/Nashville News, Clevelanders, Baltimore News \\ 
\addlinespace
\multicolumn{2}{l}{\textbf{Others}} \\
\midrule
Business & Venture Capitalists/strategists, Crypto, Law, Real Estate, Traders/Investors, Business/Entrepreneurship, Generic Brands, Farming/Agriculture, International Professionals, Home Depot \\ 
Business \& (Tech, Health, Lifestyle, or Sports) & Tech (primarily VMWare), Designers/Design Software, Salesforce, Human Resources/Human Resources Tech/Recruiters, Health Corporations, Fashion, Baseball Coaching/Personal Development \\ 
Business \& Bot & Business/Financial Freedom/Scams, Digital Marketing / Bot \\ 
Sports & Ultimate Fighting Championship, Cycling, Pro Poker, Sports Insider Accounts, Golf (Turf?), Fishing, World Wrestling Entertainment Pro Wrestlers, Supercross, Horse Racing Enthusiasts, Soccer, Wrestling, NASCAR Drivers/Analysts, Pro Golf, Professional Runners, Hockey, Women’s Basketball, College Basketball, College Football Coaches, Sprint Car \\ 
Local \& News \& Sports & Seattle News/Sports, South Carolina News/Sports, Iowa News/Sports, Kentucky News/Sports \\ 
Local \& Sports & New Orleans Media \& Sports, Auburn/‘Bama/Southeastern Conference Sports, University of Tennessee Sports, Alaska Sports, Minnesota Sports, Buffalo Sports, University of Nebraska Sports \\ 
Local \& News \& Religious \& Sports & Mormon, Utah Media, Brigham Young University Sports \\ 
News & Journalism/News about News, \red{Venezuelan Journalists}, \red{Hispanic TV Hosts/Actors} \\ 
Education (\& Business) & Administrative K–12 Education, Innovators in Education, Math Ed \\ 
Academia (\& Health) & Digital Humanities, Biologists/Geneticists, Emergency Medicine \\ 
Health & Dietitians/Nutritionists \\ 
Tech & Cybersecurity \\ 
Religious (\& Books/Comics) & Christian Leaders, Catholics,  White Christian Women Authors \\ 
Porn (\& Bot) & Porn / Bots, Adult Movie Stars, Porn \\ 
Lifestyle & Food, Weddings, Wine, Interior Design/Furnishings \\  
Celeb/Influencer & Late Millennial Celebs, International Famous People, Generic Influencers, Black Mostly-Female Celebs, Black Influencers \\ 
Entertainment & Outlander Series Enthusiast, Children’s Authors/Illustrators \\ 
Entertainment (Books/Comics) & Literary Publishers, Comics, Book Agents/Self-Promo Authors, Romance Novelists, Authors \& Publishers, Writers \\ 
Entertainment (Celeb/Influencer) & Mommy Bloggers/Brand Reps, Casting Directors/Agencies, Mainstream Content, Generic Entertainment \\ 
Entertainment (Games) & Gamers/Esports, Gaming, Magic: The Gathering, Gamers \& Content Creators \\ 
Entertainment (Music) & Electronic Music, Alt Rock, Contemporary Ensembles, Country Artists, Gospel, Jazz, K-Pop, Music Producers \\ 
Entertainment (Arts) & Creatives, Art Companies, Crafting, Concept Artists, Stand-up Comedians \\
\red{No Clear Pattern} & \red{NA / Bot, NA/Non-US, None} \\
\bottomrule
\caption{All longitudinal publics identified (right column) separated by the high-level types we applied to them during our qualitative analysis.}
\label{tab:full_categories}
\end{longtable}

\end{center}

Table~\ref{tab:full_categories} shows the full set of longitudinal publics and their high-level type. Longitudinal publics in red in Table~\ref{tab:full_categories} were excluded from the analysis of discursive overlap, either because we could not discern a clear type or because their content was primarily in a language other than English, and as such our keyword lists did not apply.

\section{Robustness of Twitter Longitudinal Publics Over Time}

\begin{figure}[!t]
    \centering
    \includegraphics[width=.8\textwidth]{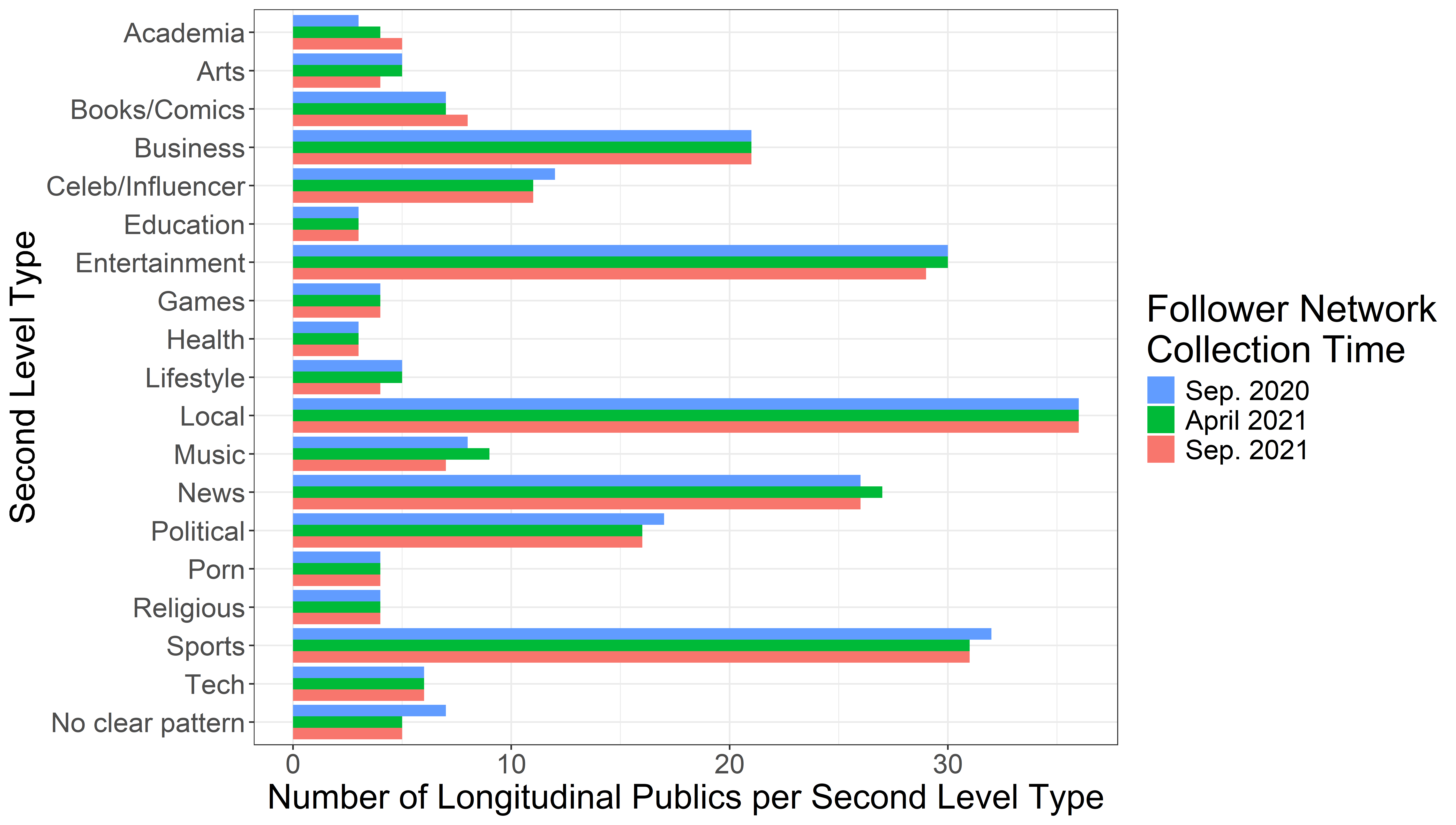}
    \caption{Share of longitudinal publics assigned to each second level type in September 2020, half a year and one year later.}
    \label{fig:clus_robustime}
\end{figure}

Figure ~\ref{fig:clus_robustime} shows how many longitudinal publics were assigned to each second level type. Each public was assigned to up to four types. On this second level we observe even more stable patterns than with the already stable composition of top creator accounts within clusters.

\section{Methods for RQ2}

\subsection{Model Specification for RQ2b} \label{sec:2bmethods}

We begin with a base model with only an intercept for issue public and then iteratively add factors for $t$ (as a factor, one per week pre/post an event), which event is being modeled, and the high-level type of longitudinal public (see below). The model that best explains the data involves a separate term for each combination of week, event, issue public, and type of longitudinal public. The full model best explains the degree of intersection between longitudinal public discourse and issue public discourse, showing statistically significantly better fit to the data using the standard ANOVA-style test ($p < 0.0001$ compared to all other models) for nested model construction compared to any subset of possible nested models. It is not only the best in comparison to the models we consider, it overall explains roughly 90\% of all variation in our outcome variable. Results below are thus presented using parameter estimates from this model.

\subsection{Model Specification for RQ2c} \label{sec:2cmethods}

In this case, our base model is a null model with fixed effects for longitudinal public (n=141) and issue public (n=2). Given the strong correlation between engagement with issue publics for longitudinal publics (see Figure~\ref{fig:rq2_overview}), we use separate fixed effects for longitudinal public and issue public, rather than a fixed effect for each issue public for each longitudinal public. 
On top of this base model, we then iteratively add factors that incorporate factors associated with individual content creators. All variables are scaled by two standard deviations and centered around their public-level mean. Our iterative model building proceeds in three steps:
\begin{enumerate}
    \item We first add variables relevant to content creator status: whether or not they are verified, and their logged ratio of followers to friends, a common measure of status \citep{joseph2021network}. 
    \item We then iteratively add demographic variables informed by exploratory analyses on which have strong correlations with the estimand of interest. We first add average audience age, then percentage of audience that is Democrat relative to Republican, then percentage of audience that is male relative to female. 
    \item Finally, we construct a network of content creators where the edge weight between content creators $A$ and $B$ is equal to the reciprocal of the size of $A$ and $B$'s shared audience. We then compute the approximate current flow betweenness centrality for each content creator in this network, providing a notion of creator centrality \emph{within} a given longitudinal public \citep{brandes2005centrality}.
\end{enumerate}
As we add each factor to the model, we incorporate interaction terms that allow for effects to vary across types of longitudinal public and/or issue public, if these interaction effects significantly improve model fit. In our first analysis, we look at which of these factors are associated with a content creator having high discursive overlap after controlling for fixed effects for both longitudinal and issue publics. Table~\ref{tab:regression_rq2c} presents results from the initial base model compared to the full model we obtain after this iterative model building process.  Finally, we also incorporate unscaled, uncentered measures of content creator audiences to assess how much variance in the outcome we can explain by creator-level factors above and beyond the fixed effects; details are provided in the Results section.

\section{Full Regression Result for RQ2b}

\footnotesize
\centering
\begin{longtable}{lll}
\hline
& Base & Full \\ \hline
(Intercept)                             & \num{-1.144}**                & \num{-4.232}**                \\
& [\num{-1.251}, \num{-1.038}] & [\num{-6.436}, \num{-2.028}] \\
issuepublicblm                          & \num{-1.955}**                & \num{1.529}*                  \\
& [\num{-2.131}, \num{-1.780}] & [\num{0.218}, \num{2.839}]   \\
metaNationalPolitics                  & \num{1.214}**                 & \num{0.368}                   \\
& [\num{1.048}, \num{1.381}]   & [\num{-0.066}, \num{0.801}]  \\
metaOthers                              & \num{-0.914}**                & \num{-0.866}**                \\
& [\num{-1.169}, \num{-0.659}] & [\num{-1.057}, \num{-0.675}] \\
issuepublicblm × metaNationalPolitics & \num{-0.917}**                & \num{-0.629}**                \\
& [\num{-1.154}, \num{-0.680}] & [\num{-0.952}, \num{-0.305}] \\
issuepublicblm × metaOthers             & \num{-0.063}                  & \num{-0.165}                  \\
& [\num{-0.270}, \num{0.144}]  & [\num{-0.361}, \num{0.032}]  \\
age                                     &                                & \num{0.060}**                 \\
&                                & [\num{0.025}, \num{0.095}]   \\
poly(par\_diff, 2)1                    &                                & \num{-1.540}*                 \\
&                                & [\num{-2.808}, \num{-0.273}] \\
poly(par\_diff, 2)2                    &                                & \num{2.755}*                  \\
&                                & [\num{0.402}, \num{5.108}]   \\
male                                    &                                & \num{-0.425}                  \\
&                                & [\num{-1.681}, \num{0.831}]  \\
log\_n                                 &                                & \num{0.104}*                  \\
&                                & [\num{0.028}, \num{0.180}]   \\
issuepublicblm × age                    &                                & \num{-0.071}**                \\
&                                & [\num{-0.090}, \num{-0.052}] \\
issuepublicblm × poly(par\_diff, 2)1   &                                & \num{0.938}*                  \\
&                                & [\num{0.069}, \num{1.807}]   \\
issuepublicblm × poly(par\_diff, 2)2   &                                & \num{1.950}*                  \\
&                                & [\num{0.582}, \num{3.317}]   \\
issuepublicblm × male                   &                                & \num{-0.508}                  \\
&                                & [\num{-1.136}, \num{0.121}]  \\
issuepublicblm × log\_n                &                                & \num{-0.041}                  \\
&                                & [\num{-0.103}, \num{0.020}]  \\
Num.Obs.                                & \num{282}                     & \num{282}                     \\
AIC                                     & \num{1928786.2}               & \num{1340872.7}               \\
BIC                                     & \num{1928808.1}               & \num{1340931.0}               \\
Log.Lik.                                & \num{-964387.109}             & \num{-670420.351}             \\
RMSE                                    & \num{0.05}                    & \num{0.04}                    \\
Std.Errors                              & by: grp                        & by: grp                        \\
\hline
\caption{Regression model tables for RQ2b. All models are linear mixed effects models with a random intercept for longitudinal public. Not shown are coefficients for the time period (week before/after event) by issue public (BLM or Electoral) by event (George Floyd or Election Day 2020) interaction terms that are included in the base model and all subsequent models.}
\label{tab:regression_rq2b}
\end{longtable}

\section{Full Regression Result for RQ2c}
\small
\centering
\begin{longtable}[h!]{p{6.1cm}rl}

\label{tab:regression_rq2c} \\

\midrule \midrule
Dependent Variable: & \multicolumn{2}{c}{Proportion of Discourse Overlap} \\
Model: & (1) & (2) \\
\midrule

\endfirsthead

\midrule \midrule
Dependent Variable: & \multicolumn{2}{c}{Proportion of Discourse Overlap} \\
Model: & (1) & (2) \\
\midrule

\endhead

\midrule
\multicolumn{3}{r}{\emph{(continued on next page)}}\\
\endfoot

\midrule \midrule

\endlastfoot

\emph{Variables} \\
IsVerified & \cellcolor{gray!20} & 0.1254$^{*}$ [0.0780; 0.1727] \\
Log(followers/following) & \cellcolor{gray!20} & -0.2642$^{*}$ [-0.2940; -0.2345] \\
Avg.Age & \cellcolor{gray!20} & 0.4610$^{*}$ [0.4396; 0.4823] \\
Prop.Dem-Prop.Rep & \cellcolor{gray!20} & -217.0$^{*}$ [-217.8; -216.3] \\
(Prop.Dem-Prop.Rep)\^2 & \cellcolor{gray!20} & 238.2$^{*}$ [236.4; 240.0] \\
Prop.Male-Prop.Female & \cellcolor{gray!20} & -0.1833$^{*}$ [-0.1920; -0.1747] \\
Approx.Bet.Cent. & \cellcolor{gray!20} & -0.3203$^{*}$ [-0.3255; -0.3150] \\
IsVerified $\times$ BLM\_IP & \cellcolor{gray!20} & -0.2460$^{*}$ [-0.2749; -0.2172] \\
Log(followers/following) $\times$ BLM\_IP & \cellcolor{gray!20} & 0.4907$^{*}$ [0.4709; 0.5105] \\
IsVerified $\times$ Nat.Politics\_LP & \cellcolor{gray!20} & -0.1327 [-0.3491; 0.0837] \\
IsVerified $\times$ Other\_LP & \cellcolor{gray!20} & 0.1699$^{*}$ [0.1000; 0.2398] \\
Log(followers/following) $\times$ Nat.Politics\_LP & \cellcolor{gray!20} & 0.2361$^{*}$ [0.1826; 0.2896] \\
Log(followers/following) $\times$ Other\_LP & \cellcolor{gray!20} & -0.0164 [-0.0593; 0.0266] \\
BLM\_IP $\times$ Nat.Politics\_LP & \cellcolor{gray!20} & -0.9809$^{*}$ [-1.027; -0.9347] \\
BLM\_IP $\times$ Other\_LP & \cellcolor{gray!20} & -0.1162$^{*}$ [-0.1438; -0.0886] \\
Nat.Politics\_LP $\times$ Avg.Age & \cellcolor{gray!20} & -0.2128$^{*}$ [-0.2742; -0.1514] \\
Other\_LP $\times$ Avg.Age & \cellcolor{gray!20} & -0.1099$^{*}$ [-0.1101; -0.1097] \\
BLM\_IP $\times$ Avg.Age & \cellcolor{gray!20} & -0.5634$^{*}$ [-0.5649; -0.5620] \\
Nat.Politics\_LP $\times$ Prop.Dem-Prop.Rep & \cellcolor{gray!20} & 181.3$^{*}$ [174.4; 188.1] \\
Other\_LP $\times$ Prop.Dem-Prop.Rep & \cellcolor{gray!20} & 55.95$^{*}$ [53.66; 58.23] \\
Nat.Politics\_LP $\times$ (Prop.Dem-Prop.Rep)$^2$ & \cellcolor{gray!20} & -245.4$^{*}$ [-245.7; -245.1] \\
Other\_LP $\times$ (Prop.Dem-Prop.Rep)$^2$ & \cellcolor{gray!20} & -193.9$^{*}$ [-197.3; -190.6] \\
BLM\_IP $\times$ Prop.Dem-Prop.Rep & \cellcolor{gray!20} & -41.35$^{*}$ [-44.40; -38.30] \\
BLM\_IP $\times$ (Prop.Dem-Prop.Rep)$^2$ & \cellcolor{gray!20} & -80.38$^{*}$ [-84.14; -76.61] \\
Nat.Politics\_LP $\times$ Prop.Male-Prop.Female & \cellcolor{gray!20} & 0.2171$^{*}$ [0.2079; 0.2262] \\
Other\_LP $\times$ Prop.Male-Prop.Female & \cellcolor{gray!20} & 0.1405$^{*}$ [0.1344; 0.1466] \\
BLM\_IP $\times$ Prop.Male-Prop.Female & \cellcolor{gray!20} & -0.2915$^{*}$ [-0.3047; -0.2782] \\
Nat.Politics\_LP $\times$ Approx.Bet.Cent. & \cellcolor{gray!20} & 0.2382$^{*}$ [0.2353; 0.2412] \\
Other\_LP $\times$ Approx.Bet.Cent. & \cellcolor{gray!20} & 0.2386$^{*}$ [0.2320; 0.2453] \\
BLM\_IP $\times$ Approx.Bet.Cent. & \cellcolor{gray!20} & 0.1945$^{*}$ [0.1927; 0.1964] \\
IsVerified $\times$ BLM\_IP $\times$ Nat.Politics\_LP & \cellcolor{gray!20} & 0.4436$^{*}$ [0.4192; 0.4681] \\
IsVerified $\times$ BLM\_IP $\times$ Other\_LP & \cellcolor{gray!20} & 0.2456$^{*}$ [0.2336; 0.2577] \\
Log(followers/following) $\times$ BLM\_IP $\times$ Nat.Politics\_LP & \cellcolor{gray!20} & -0.5237$^{*}$ [-0.5496; -0.4978] \\
Log(followers/following) $\times$ BLM\_IP $\times$ Other\_LP & \cellcolor{gray!20} & -0.6495$^{*}$ [-0.6716; -0.6274] \\
BLM\_IP $\times$ Nat.Politics\_LP $\times$ Avg.Age & \cellcolor{gray!20} & 0.3566$^{*}$ [0.3538; 0.3594] \\
BLM\_IP $\times$ Other\_LP $\times$ Avg.Age & \cellcolor{gray!20} & 0.0625$^{*}$ [0.0599; 0.0651] \\
BLM\_IP $\times$ Nat.Politics\_LP $\times$ Prop.Dem-Prop.Rep & \cellcolor{gray!20} & 66.30$^{*}$ [61.17; 71.42] \\
BLM\_IP $\times$ Other\_LP $\times$ Prop.Dem-Prop.Rep & \cellcolor{gray!20} & 14.14$^{*}$ [7.904; 20.38] \\
BLM\_IP $\times$ Nat.Politics\_LP $\times$ (Prop.Dem-Prop.Rep)$^2$ & \cellcolor{gray!20} & 70.39$^{*}$ [65.43; 75.35] \\
BLM\_IP $\times$ Other\_LP $\times$ (Prop.Dem-Prop.Rep)$^2$ & \cellcolor{gray!20} & 68.69$^{*}$ [65.69; 71.69] \\
BLM\_IP $\times$ Nat.Politics\_LP $\times$ Prop.Male-Prop.Female & \cellcolor{gray!20} & 0.2134$^{*}$ [0.2029; 0.2239] \\
BLM\_IP $\times$ Other\_LP $\times$ Prop.Male-Prop.Female & \cellcolor{gray!20} & 0.0772$^{*}$ [0.0661; 0.0883] \\
BLM\_IP $\times$ Nat.Politics\_LP $\times$ Approx.Bet.Cent. & \cellcolor{gray!20} & -0.1649$^{*}$ [-0.1660; -0.1638] \\
BLM\_IP $\times$ Other\_LP $\times$ Approx.Bet.Cent. & \cellcolor{gray!20} & -0.1603$^{*}$ [-0.1697; -0.1509] \\

\midrule
\emph{Fixed-effects} \\
Issue Public Type & Yes & Yes \\
Longitudinal Public ID & Yes & Yes \\

\midrule
\emph{Fit statistics} \\
Observations & 515{,}756 & 515{,}756 \\
Squared Correlation & 0.49216 & 0.54255 \\
Pseudo R$^2$ & 0.88913 & 0.90149 \\
BIC & 6{,}559{,}174.5 & 5{,}828{,}508.8 \\

\midrule
\multicolumn{3}{l}{\emph{Signif. Codes: *: 0.001}} \\
\caption{Regression model tables for RQ2c. This table compares two fixed effects modeling strategies, namely the base model describe above and the full model that is iteratively constructed by adding fixed effects related to creator and audience characteristics.}
\end{longtable}

\end{appendices}

\bibliography{interactapasample, leftovers}%

\end{document}